\documentclass[usegraphicx,usenatbib]{mn2e}

\newcommand{\km}{\,\mbox{km}\,\mbox{s}^{-1}}

\title[3D spectroscopy of  Mrk 334]{3D spectroscopy of  merger Seyfert galaxy Mrk 334: nuclear starburst, superwind and the circumnuclear cavern}

\author[A.   Smirnova  and A.   Moiseev]{Aleksandrina Smirnova$^1$\thanks{ssmirnova@gmail.com} and Alexei Moiseev$^1$\\
$^1$Special Astrophysical Observatory, Russian Academy of Sciences, Nizhnii Arkhyz,  Karachaevo-Cherkesskaya Republic,
369167 Russia}

\begin{document}

\maketitle

\begin{abstract}

We are presenting  new results on kinematics and structure of the Mrk 334 Seyfert galaxy. Panoramic (3D) spectroscopy is performed at the 6-m telescope of the Special Astrophysical Observatory of the Russian Academy of Sciences using the  MPFS integral-field spectrograph and scanning Fabry--P\'{e}rot interferometer. The deep images have revealed that Mrk 334 is observed during the final stage of its merging with a massive companion. A possible mass ratio ranges from $1/5$ to $1/3$. The merger has triggered mass redistribution in the disk  resulting in an intensification of nuclear activity and in a burst of star formation in the inner region of the galaxy. The circumnuclear starburst is so intense that its contribution to the  gas ionization exceeds that contribution of the AGN. We interpret the nuclear gas outflow  with velocities of $\sim200\km$ as a galactic superwind that accompanies the violent star formation. This suggestion is consistent with the asymmetric X-ray brightness distribution in Mrk 334. The trajectory of the fragments of the disrupted satellite  in the vicinity of the main galaxy nucleus can be traced. In the galaxy disk a cavern is found that is filled with a low-density ionized gas. We consider this region to be  the place where the remnants of  the companion  have recently penetrated through the gaseous disk of the main galaxy.
\end{abstract}


\section{Introduction}
\label{sec0}
Numerical simulations demonstrate that galaxy interaction stimulates a concentration of gas in its central regions, thereby triggering nuclear activity and/or a burst of star formation \citep*{BarnesHernquist1991, Springel2005, Bekki2006}.
Many authors have tried to find a correlation between  an AGN phenomenon and galaxy environment: the presence of companions or traces of interaction \citep*{Dahari1985,DeRobertis1998, Schmitt2001, Knapen2005}. However, statistically significant correlation has not been found.

A number of authors suggest that the activity may be triggered and sustained by a complex mechanism that includes several factors (see \citet{Martini2004} and the references therein). It is clear that only a detailed analysis of the kinematics and dynamics of both the inner  ($100-1000$~pc  scale) and outer regions in active galaxies would make it possible to understand how in each particular case the  `fuel' (interstellar gas) is brought into the domain of action of the gravitational forces of the  AGN `central engine'.

This paper continues a series of papers dedicated to a detailed study of the inner kinematics of
active galaxies via   methods of  panoramic (3D) spectroscopy. This technique  provides spectra for every spatial element (`spaxel') of a two-dimensional field of view. It is a powerful tool for studying non-circular motions and  gas ionization properties both in the circumnuclear and external regions.
Our work is aimed to investigate the relation between the gas kinematics, morphological features and nuclear activity in individual galaxies as well as the mechanisms of the central region feeding. We have already published the results concerning Mrk 315 \citep{Ciroi2005}, NGC 6104 \citep*{Smirnova2006}, and Mrk 533 \citep{Smirnova2007}. In this paper we report a detailed study of  Mrk 334.

Mrk 334 (VV 806, UGC 6) is a peculiar galaxy with  Sy1.8 nucleus (according to the NED database). This object has been popular among the researchers  due to its peculiar appearance on optical images: it has an asymmetric eastward-extending arm
\citep{VorontsovV1977}  and a bright H$\alpha$ condensation near the nucleus \citep{GonzalezDelgado1997}. The latter authors  suggested that the object actually consisted of two merging galaxies. Mrk 334 is notable for violent star formation, resulting in high IR luminosity $L_{IR}=8.9\times10^{11}\,$L$_{\odot}$ \citep{Perez2001}.  \citet{RothbergJoseph2004} classify it as a luminous  infrared galaxy (LIRG).  \citet{Maiolino1997} suggested that the peculiarities in the
structure of the galaxy are indicative of its recent interaction with a companion, which has triggered the nuclear activity. However, when and what did the galaxy interact with? In the present paper we try  to answer this question and to look for the feedback effects between the central and surrounding regions.

The paper has the following layout. Section~\ref{sec1} describes the observations and the data reduction; Section~\ref{sec2} analyses the distribution of ionized gas and stars both in the inner disk and in the outer regions of the galaxy. Section~\ref{sec3}
analyses the ionization state of selected regions in  Mrk 334, and in Section~\ref{sec4}   the
kinematics of the gas and stars are considered. Section~\ref{sec5} studies the  peculiarities of  X-ray radiation according to the \textit{ROSAT} data, and Section~\ref{sec6} includes an overall discussion of the whole galaxy structure  and the circumnuclear ionized gas cavern.

The adopted distance to the galaxy  --- 91.4~Mpc \citep{Maiolino1997} corresponds to a scale of 443~pc/arcsec (for redshift z=0.022 and $H_0=75\km\,\mbox{Mpc}^{-1}$).

\section{Observations and Data Reduction}
\label{sec1}

All observations were made in the prime focus of the 6-m telescope of the
Special Astrophysical Observatory of the Russian Academy of Sciences (SAO RAS). Table~\ref{tnab} provides the log of the observations. The central region of Mrk 334 was observed with the Multi-Pupil Fiber Spectrograph (MPFS). The large-scale kinematics and galactic environment were studied using the SCORPIO multi-mode focal reducer operating in the modes of scanning Fabry--P\'{e}rot interferometer (FPI) and broad-band imaging. The detectors used in 2006 and 2002  were a CCD EEV42-40 ($2048\times2048$ pixels) and a CCD TK1024 ($1024\times1024$ pixels), correspondingly.

\subsection{MPFS integral-field spectrograph}

The integral-field  spectrograph  MPFS \citep{Afanasiev2001} takes simultaneous spectra of 256 spatial elements arranged in the form of  $16\times16$ square lenses  array with a scale of  1 arcsec per spaxel. Behind each lens an optical fibre is located whose other end is packed into the  spectrograph slit.
The sky background spectrum was simultaneously taken  with 17 additional fibres located  4 arcmin away from the object. The wavelength interval included numerous  emission lines of ionized gas and absorption features of the stellar population.

The preliminary data reduction steps were described  earlier  \citep{Moiseev2004,Smirnova2007}. Reduction yields a  `data cube', where each pixel in the $16\times16$ arcsec field  has a spectrum associated with it. The spectra of the spectrophotometric standard stars were used to convert counts into absolute fluxes. Observations were made successively in two spectral intervals (see Table~\ref{tnab}). The overlap of the two spectral domains  allowed us to join them so as to operate with a single data cube covering a $\lambda\lambda3740-7220$\AA spectral range.

To construct the stellar velocity field, we use the cross-correlation technique adapted for MPFS data \citep{Moiseev2001}. Spectra from MILES library \citep{Sanchez2006} smoothed to the instrumental resolution were adapted as templates for cross-correlation. We have mapped the  line-of-sight velocity and brightness distribution fields for the main emission lines using the Gaussian fitting of their profiles. Underlying absorption lines were taken in account as approximation by a linear combination of smoothed and redshifted  MILES templates.
\begin{table}
\caption{Log of the observations.} \label{tnab}
\begin{tabular}{@{}crrcrr}
\hline
    Date       &   Instrument      & Exp.       & Sp.          & Sp.     & seeing    \\
               &                   & time,  s   & range         &resol.     &    arcsec \\
\hline
23.11.2006  &   MPFS             &   7200  &$3740-5850$\,\AA    &      6.5  \AA    &    1.4   \\
              &                    &   7200  &$4300-7220$\,\AA    &      6.5  \AA    &    1.4  \\
\hline
05.09.2002  &  SCORPIO           &   6400  &     H$\alpha$    &     2.8  \AA     &    1.3       \\
              &  (FPI)             &         &                  &             &              \\
23.10.2006   & SCORPIO           &   660   &    $V$           &             &    1.4       \\
               & (Images)          &   1020  &    $R$           &             &    1.4       \\

\hline
\end{tabular}
\end{table}

\begin{figure}
\centerline{
\includegraphics[width=4.2cm]{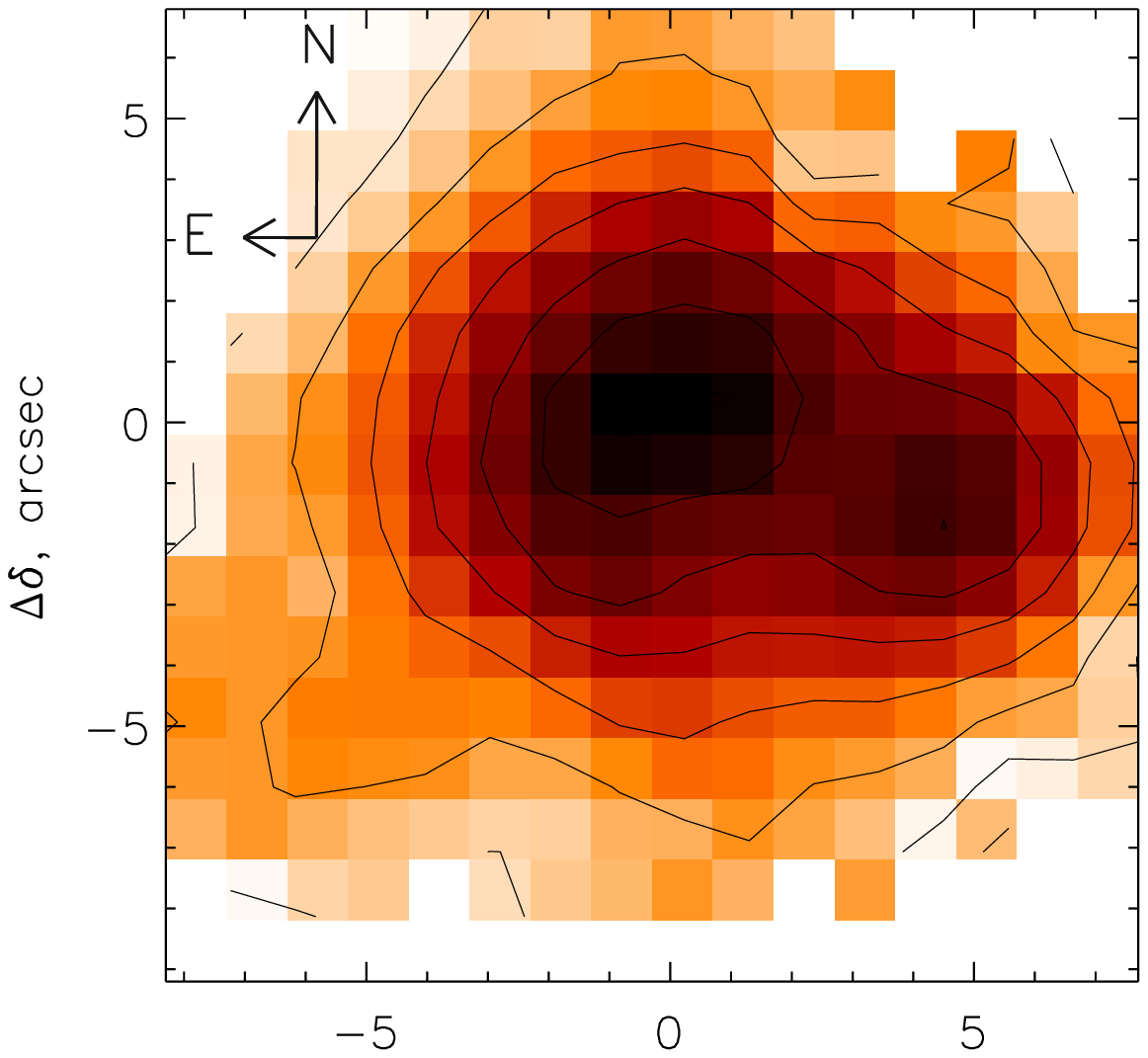}
\includegraphics[width=4.2cm]{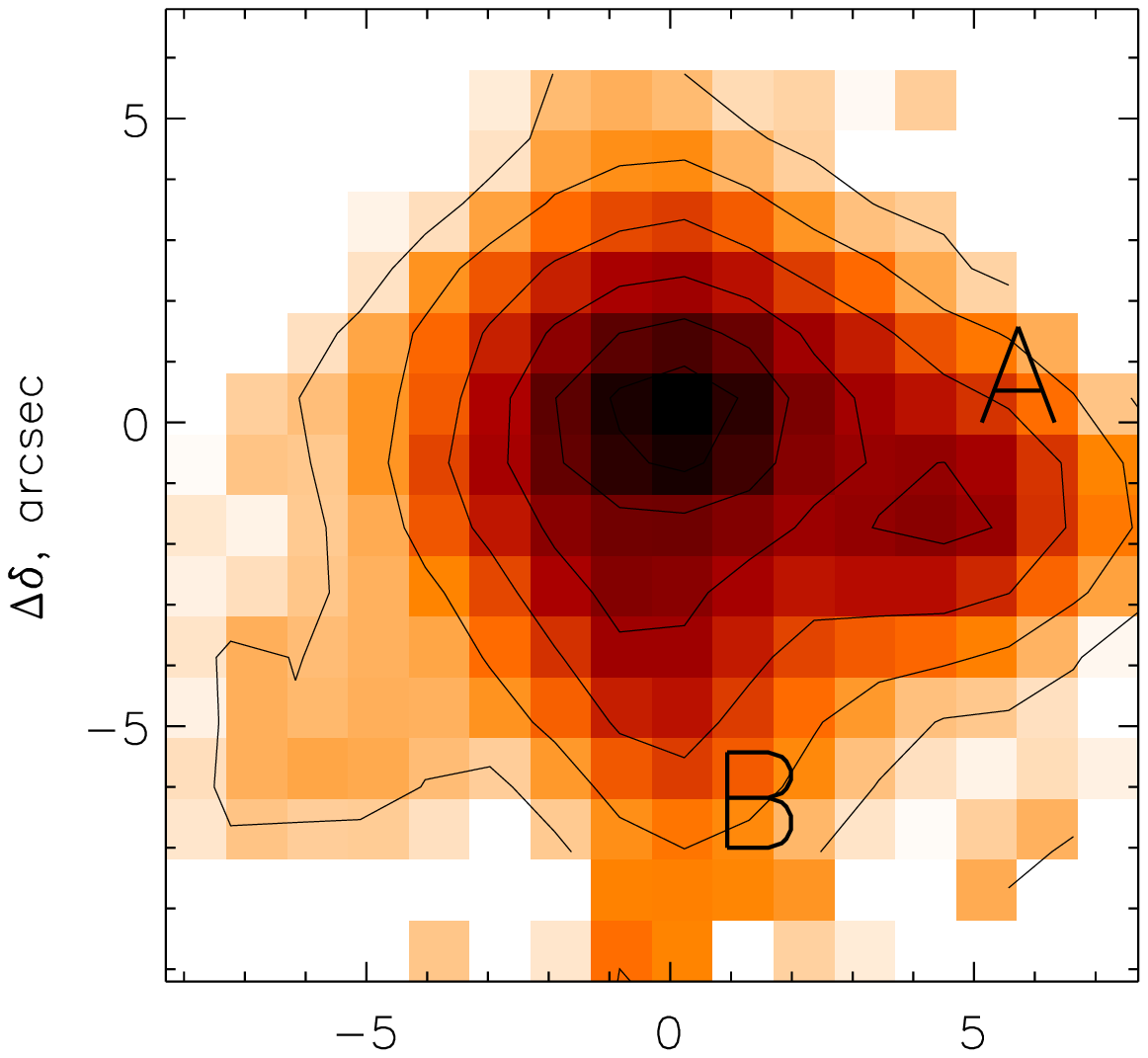}}
\centerline{
\includegraphics[width=4.2cm]{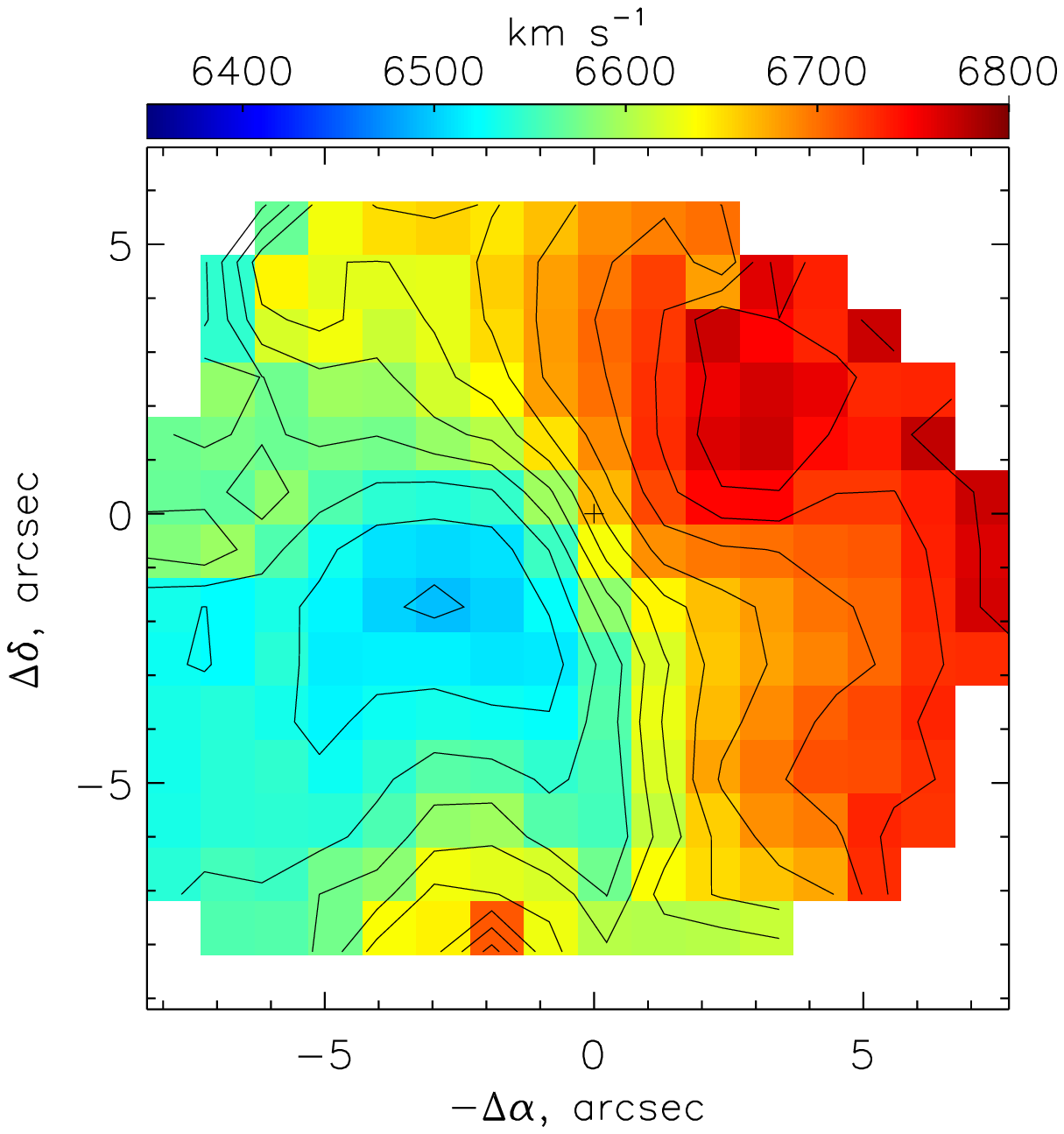}
\includegraphics[width=4.2cm]{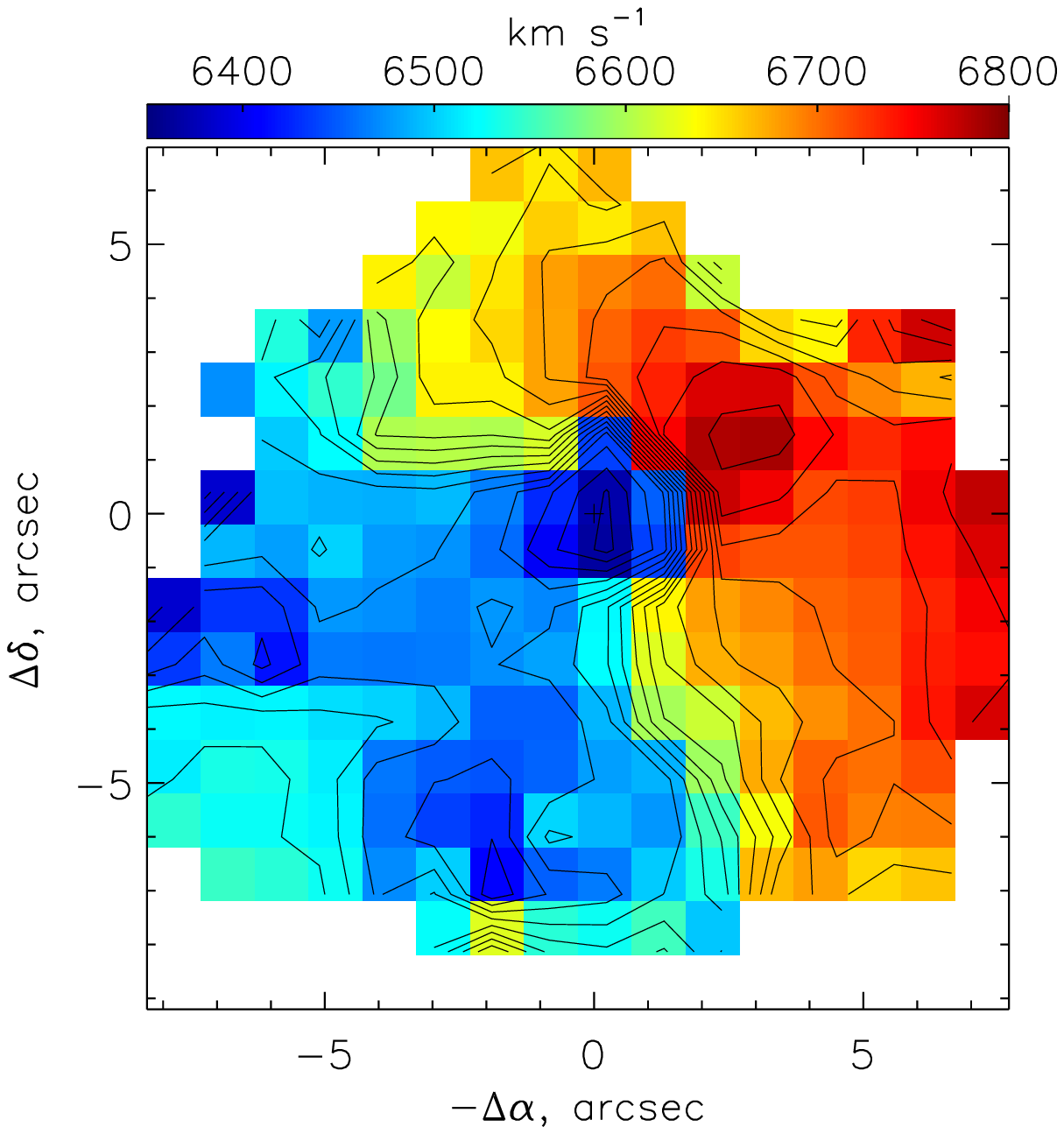}
}
\caption{Images (top) and velocity fields (bottom) in the H$\alpha$ (left) and [OIII]$\lambda\lambda4959,5007$ (right) emission lines according to MPFS data. Regions `A' and `B' are marked.}
\label{f01}
\end{figure}

\subsection{SCORPIO}

SCORPIO universal instrument \citep{AfanasievMoiseev2005} allows various spectroscopic and photometric
observations to be performed  within $6$ arcmin field of view. Below we describe
each of the modes employed in detail.

\subsubsection{Fabry--P\'{e}rot Interferometer}

We used the scanning FPI operating in the H$\alpha$ emission line to study the kinematics of the ionized gas. During the  observations we successively took 32 interference images of the object by changing the FPI plate gap.
A detailed description of the technique of observations and data reduction can be found in the
papers by \citet{Moiseev2002} and \citet{Moiseev2008}. This reduction yields a data cube, where
a 32-channel spectrum with a sampling step of 0.9~\AA\ is attached to each  $0.28$ arcsec  pixel. The spectroscopic resolution is 2.8~\AA. The velocity field of the ionized gas and images in the H$\alpha$ emission line were mapped using Gaussian fitting of the emission-line profiles.  We also generated an image of the galaxy in the continuum close to the emission line.

We calibrated the emission-line flux map into absolute energy units ($\mbox{erg}\,\mbox{s}^{-1}\,\mbox{cm}^{-2}$) by comparing it with the  H$\alpha$ distribution according to the  MPFS data for the central region. The total H$\alpha$-luminosity of the galaxy was found to be $2.3\times10^{41}\,\mbox{erg}\,\mbox{s}^{-1}$ given the adopted extinction of  $A_R=1.5^m$. The extinction estimate is based on the  H$\alpha/$H$\beta$ line intensity ratio for the HII region to the west of the nucleus (hereinafter referred to as `Region A'). According to \citet{Kennicutt1998ARA}, such a luminosity   corresponds to star-formation rate of  SFR=$18\,$M$_\odot/\mbox{yr}$, if we neglect a contribution from the AGN in the H$\alpha$ flux.

\subsubsection{Direct Images}

We took  images of the galaxy in the Johnson--Cousins {\it V} and {\it R} bands with a sampling of   0.35 arcsec per pixel.  Non-photometric weather conditions prevented the use of standard stars to calibrate fluxes.  We performed a coarse calibration in the {\it V} band  based on the aperture photometry data listed in the HyperLeda database \textrm{(http://leda.univ-lyon1.fr/)}. An accuracy of the zero-point is $0.1-0.2\,\mbox{mag}$. We calibrated the {\it R} image assuming that the average colour index in the disk of Mrk 334 corresponds to that of a normal Sb-Sc-type galaxy: {\it (V-R)}$\approx0.5$. The assumption of the normal $V-R$ color is justified by the known normal 2MASS infrared colours of Mrk 334. The depth of the surface-brightness measurements reaches $25.5\,\mbox{mag}\,\mbox{arcsec}^{-2}$ in the {\it V} and {\it R} bands, which is significantly deeper  than the other known images of Mrk 334.

\begin{figure*}
\centerline{
\includegraphics[width=8cm]{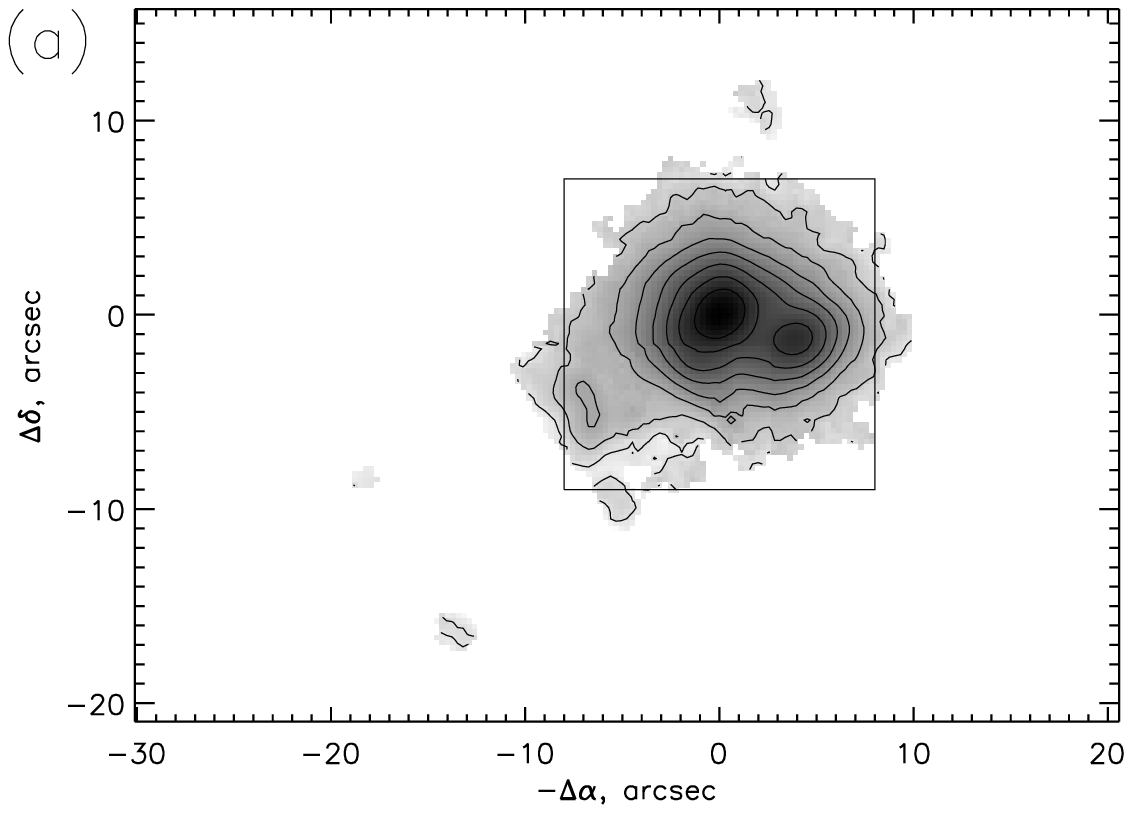}
\includegraphics[width=8cm]{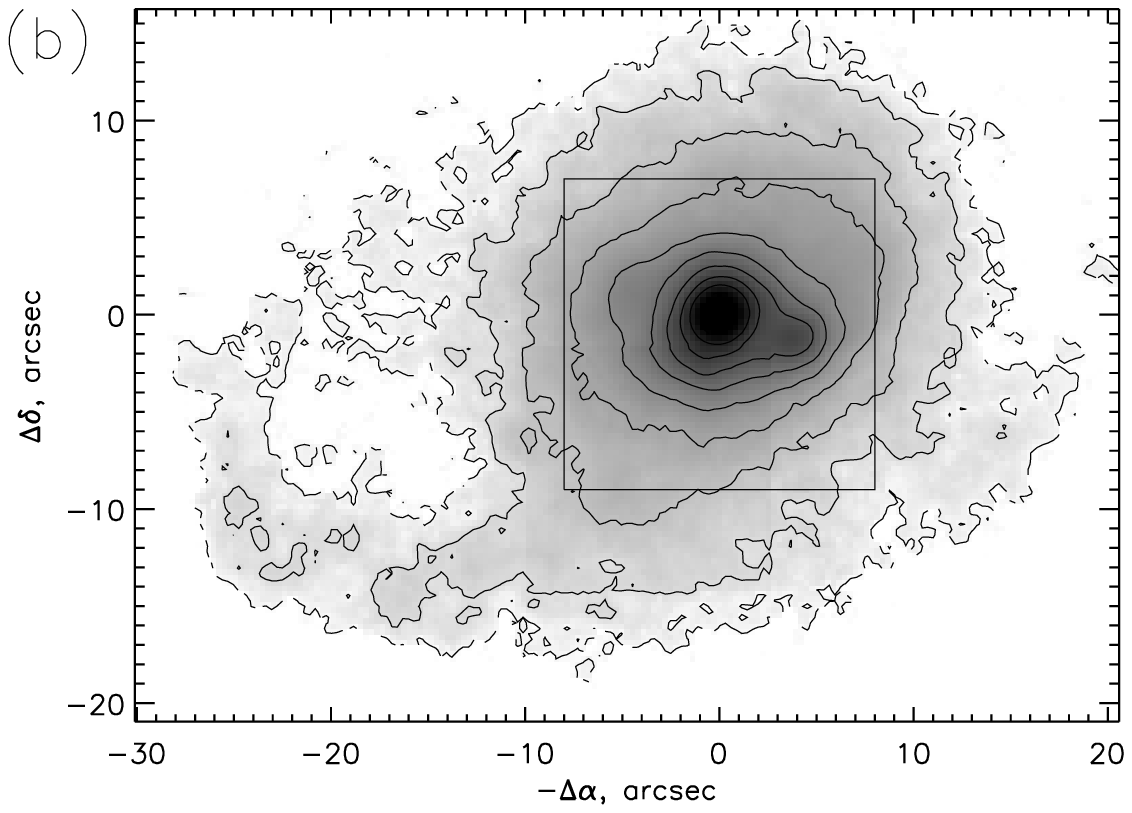}}
 \caption{FPI images of Mrk 334 in the H$\alpha$ line (a) and in the continuum
(b). The square indicates the  region observed with the MPFS}\label{f02}
\end{figure*}

\begin{figure*}
\includegraphics[width=8cm]{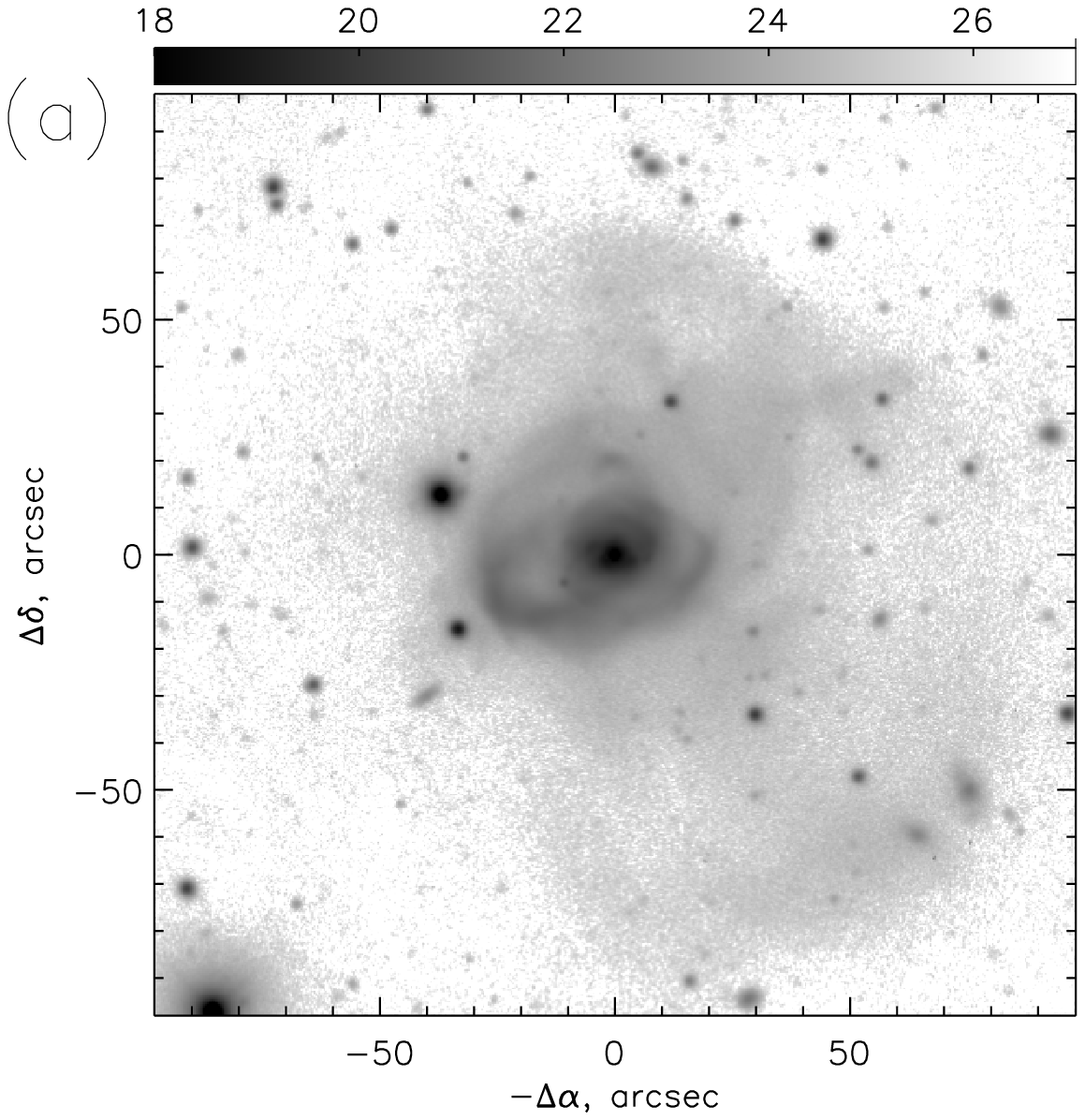}
\includegraphics[width=8cm]{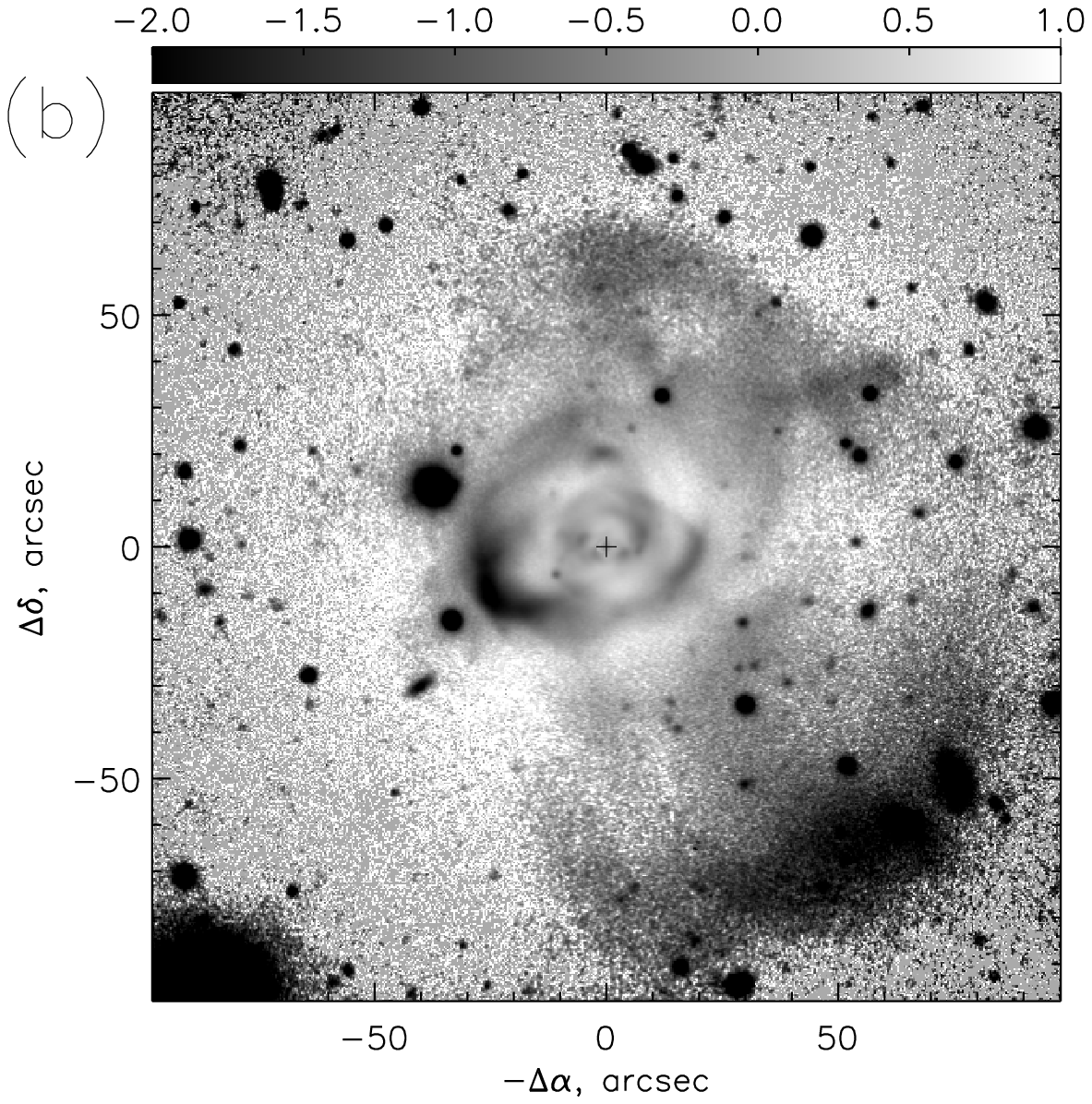}
\caption{$R$-band image of the galaxy taken with the 6-m telescope (a) and a
residual image after the subtraction of the two-dimensional model (b). The scale is in $\mbox{mag}\,\mbox{arcsec}^{-2}$.
}\label{f03}
\end{figure*}

\section{Morphological features of Mrk 334}
\label{sec2}
\subsection{Line and Continuum Images}

We use the  MPFS spectra to construct the maps in various emission lines covering the $7\times7$~kpc  central region of  Mrk 334 (see Fig.~\ref{f01}). The maps show some other bright regions besides the nucleus. The brightest of them, the Region `A', is located 4 arcsec to the west of the nucleus. \citet{GonzalezDelgado1997} were the first
to find it in H$\alpha$. At our maps it can be seen in other emission lines
and in the continuum. Fainter Region `B' is located  $r=3-4$ arcsec to the south of the nucleus and shows up  mostly only in the [OIII]$\lambda\lambda4959,5007$ doublet. The continuum image exhibits,
in addition to Region `A', an amorphous structure that we refer to as Region `C'.

Many authors (see Introduction) believed Mrk 334 to be an interacting system. Their conclusion was based mostly on the presence of an asymmetric spiral arm to the east of the nucleus resembling a tidal tail \citep{Hunt1997,RothbergJoseph2004}. This
feature is  clearly visible on the FPI continuum image and is absent in the H$\alpha$ line map (Fig.~\ref{f02}).
Therefore, this arm does not harbor star-forming regions.

\begin{table}
\caption{Parameters of the Photometric Decomposition.} \label{tab_2}
\begin{tabular}{rlll}
\hline
Component      &  parameter   & $V$          & $R$ \\
               &              & filter     & filter          \\
\hline
Sersic's bulge &   n          &   2       &  2   \\
               &   $\mu_{eff}$& $20.3$~mag   &  $19.8$~mag\\
               &   $r_0$      & 3\farcs2 (1.4 kpc) &  3\farcs2 (1.4 kpc)\\
Inner  disk    &   $\mu_0$    & $21.5$~mag   &  $20.9$~mag\\
               &   $h$        & 7\farcs8 (3.5 kpc) &  8\farcs7 (3.9 kpc)\\
Outer  disk    &   $\mu_0$    & $24.8$~mag   &  $24.2$~mag\\
               &   $h$        & $32''$ (14 kpc) &  $43''$ (19 kpc)\\
\hline
\end{tabular}
\end{table}

\begin{figure}
\includegraphics[width=8cm]{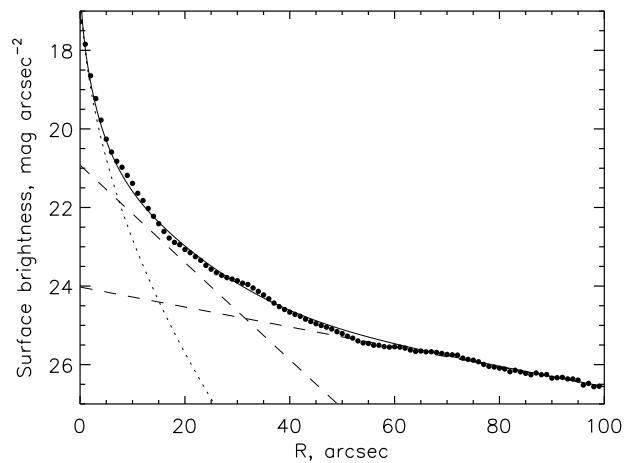}
\caption{The mean $R$-band brightness profile (the bold dots) and its model fit (the
solid line). The dashed lines show the contribution of each of the disks. The dotted line marks the bulge.
}\label{f04}
\end{figure}

\begin{figure*}
\includegraphics[width=5.5cm]{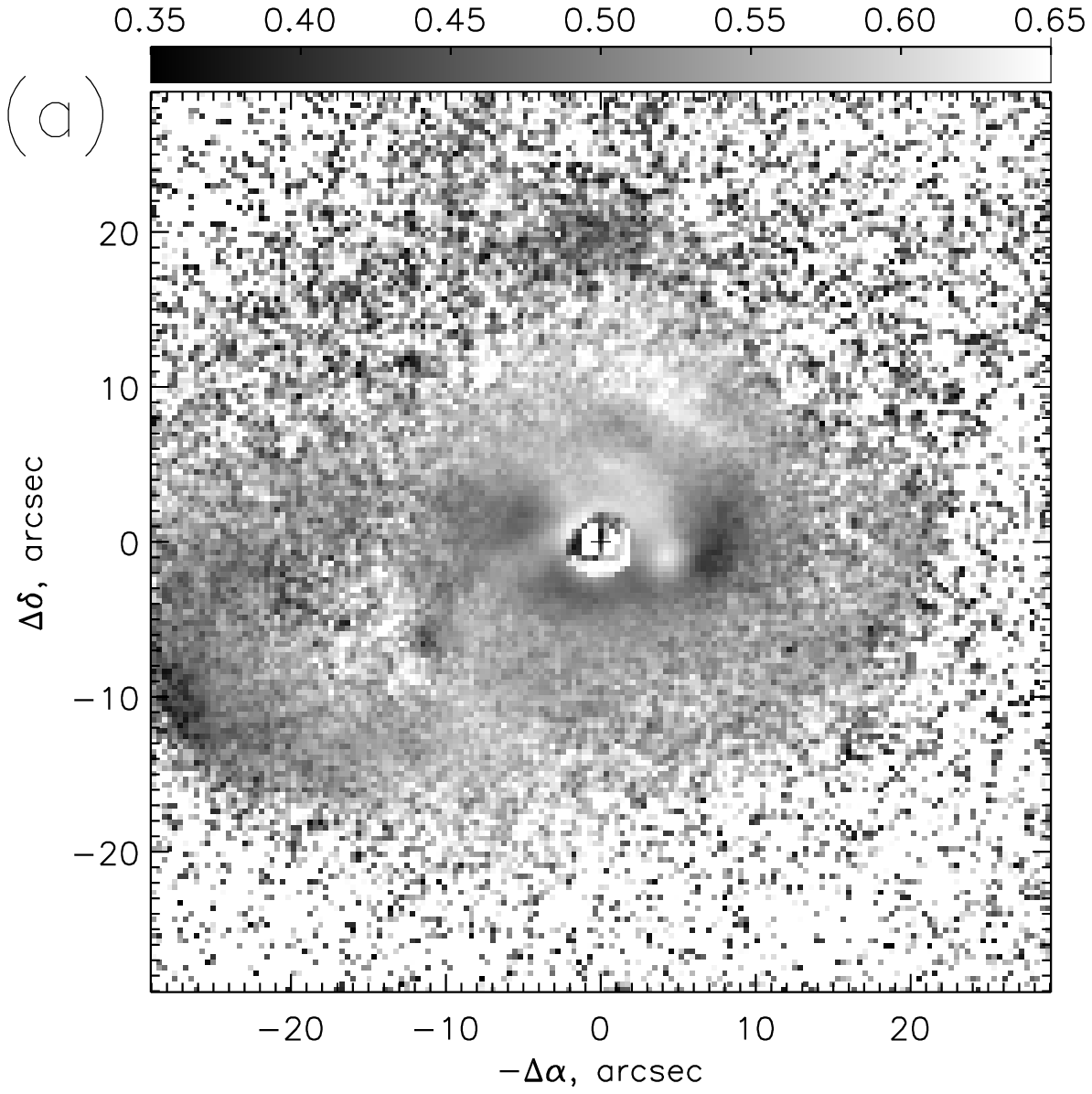}
\includegraphics[width=5.5cm]{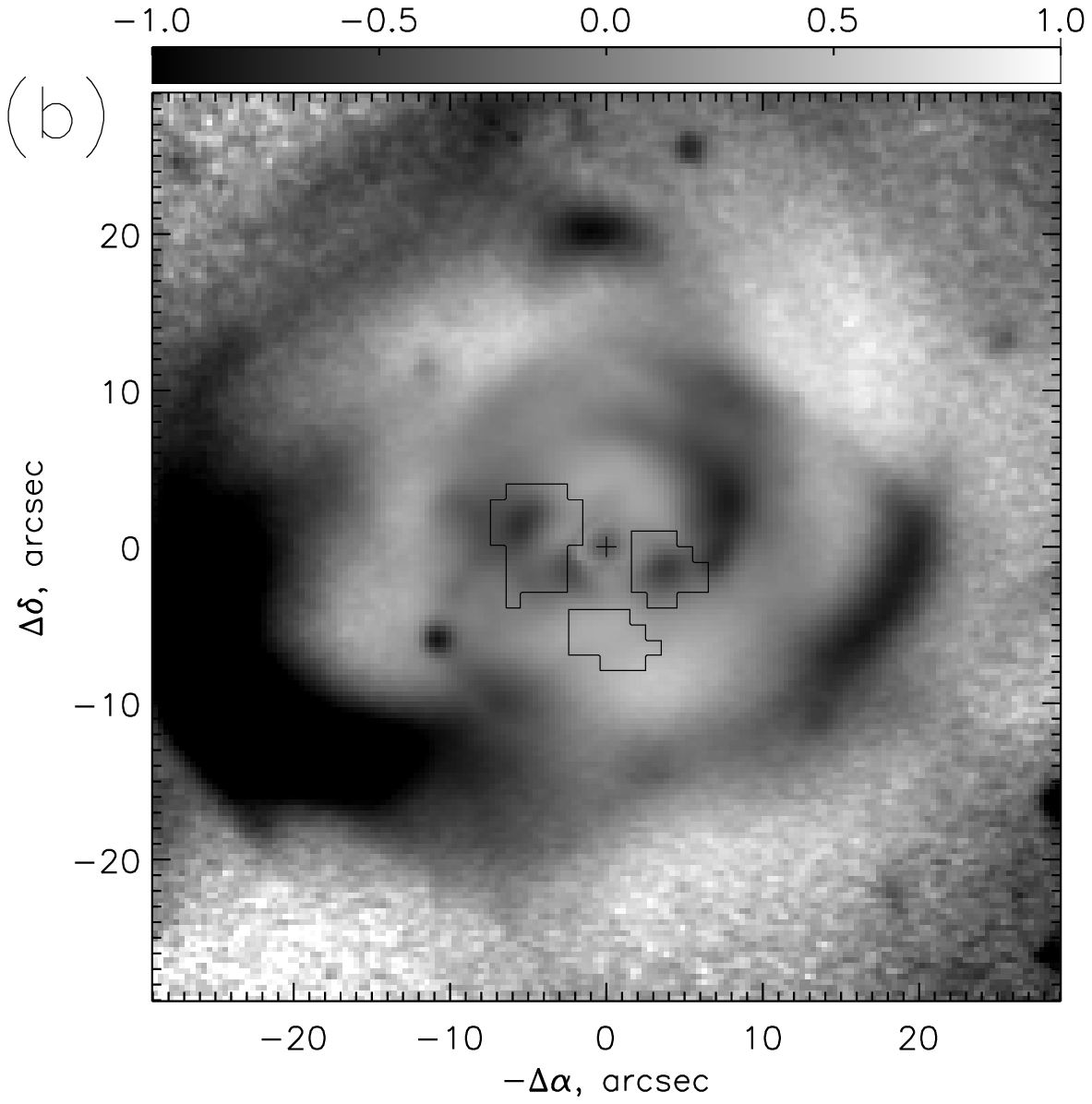}
\includegraphics[width=5.5cm]{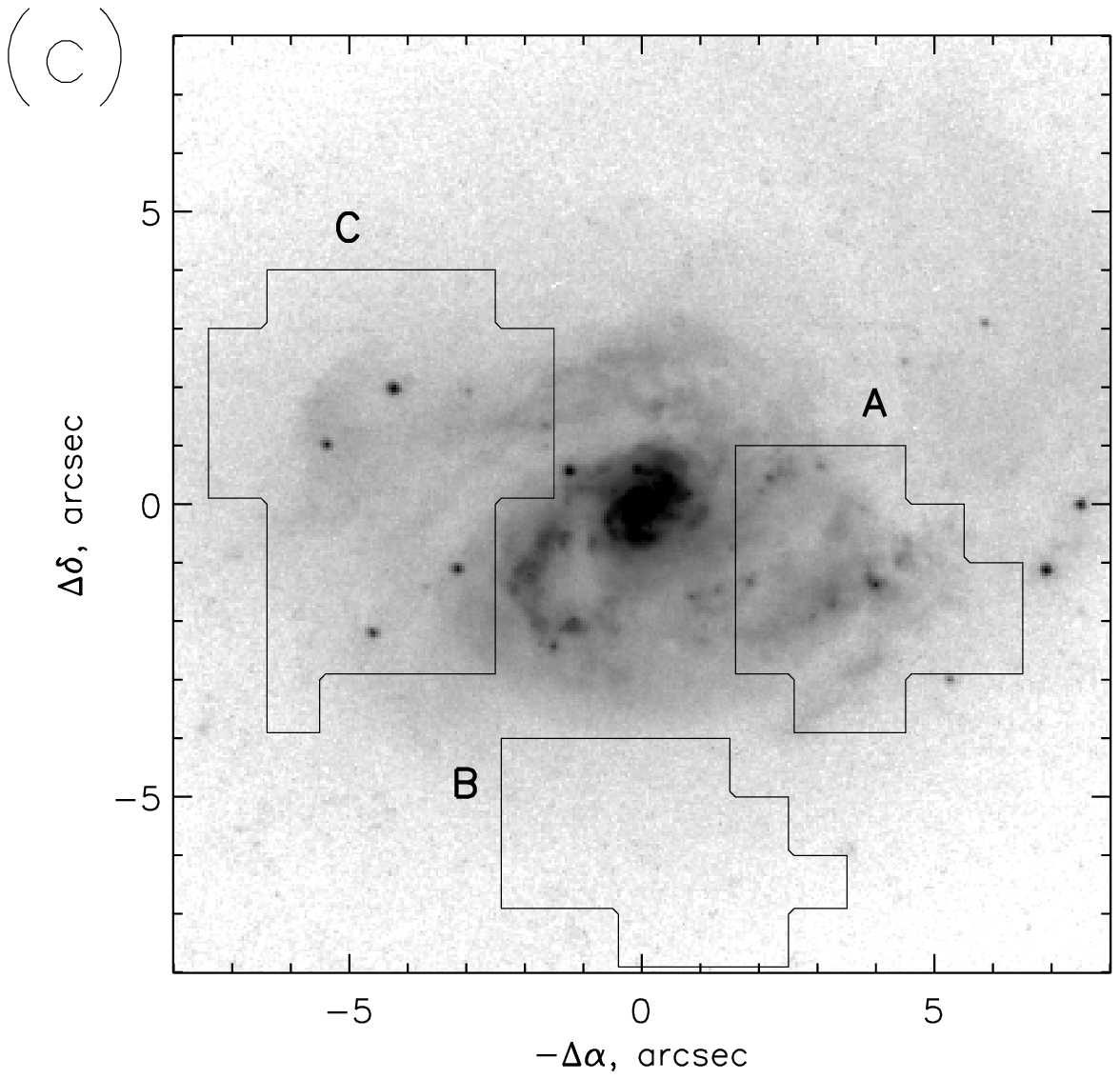}
\caption{The distribution of the $(V-R)$ colour index in the inner regions of the
galaxy (a); residual brightness distribution for the same region (b), and the
HST/WFPC2 archive image in the F606W filter (c). The contours in figures~(b) and (c) show the boundaries of the condensations identified via MPFS observations.}\label{f05}
\end{figure*}

\subsection{Multicomponent Structure of the Disk and Outer Filaments}

The deep images of the galaxy show that the  tidal
arm noted above  is the brightest part of the vast system of shells and lower surface brightness
filaments (Fig.~\ref{f03}). The shells have sharp outer edges located about  70 arcsec northwest and about  100 arcsec southwest of the nucleus, which corresponds to 31 and 44~kpc, respectively. The {\it R} band surface brightness of the outer regions is  about  $24-25\,\mbox{mag}\,\mbox{arcsec}^{-2}$. Similar arc-like features are typical of galaxies currently interacting or having interacted in the past with a companion
\citep{SchweizerSeitzer1988, WehnerGallagher2005}.

To study the brightness distribution in the filaments, we must remove the  axisymmetric components of the galaxy -- the bulge and the disk. To decompose the image into components, we use an iterative method of constructing one- and two-dimensional models \citep{Moiseev2004,Ciroi2005}. The idea of the method is to
determine the parameters of the exponential disk from the outer parts of the azimuthally averaged
brightness profile and subtract the resulting brightness distribution from the original image.
The residual image is averaged over round apertures and fitted to the S\'{e}rsic's profile for
the bulge.  We then subtract the bulge model from the initial image and use the residual image
to build the next iteration for the disk. When constructing the model we masked the high-contrast features like stars and tidal spirals.
We found the galaxy image to be best approximated by the model consisting of a bulge and
two exponential disks with different radial scales. Table~\ref{tab_2} lists the parameters of the
photometric components. Here $n,r_{e}$, and $\mu_{eff}$ are the exponent, effective radius, and brightness
of the  bulge, respectively, and $\mu_0$ and $h$ are the central surface brightness and radial
disk scalelength. The position angle and the apparent ellipticity of the disks were fixed in accordance with the orientation of the ionized-gas disk (Section~\ref{sec4}). The results of decomposition performed in the two filters agree fairly well with each other, except that
the scalelength of the outer disk is larger by a factor of 1.3 in the $R$ band.

Figure~\ref{f04} shows the surface-brightness profile computed by averaging the brightness over
elliptical rings.   It shows a well-defined break at $r=50-60$ arcsec and is dominated by the
outer disk at larger galactocentric distances. Such multicomponent (two-tiered) disks
have now become increasingly popular among the researchers. According to the classification
proposed by  \citet{Erwin2005},  Mrk 334 exhibits a  typical type III (antitruncated)
surface-brightness profile. Among the galaxies studied by \citet{Erwin2005} such profile
characterizes mostly post-interacting objects. A detailed study of individual galaxies also suggests that interaction events may result in the formation of multi-tiered disks. Examples include NGC 615 \citep{Silchenko2001}, Mrk 315 \citep{Ciroi2005}, NGC 7217 and NGC 7742 \citep{SilchenkoMoiseev2006}.

Mrk 334 appears to represent configuration, where the debris of the companion torn apart by the tidal forces precess in the plane of the galaxy. The outer disk is being formed right now with a relatively long scalelength and a
low central brightness. Indeed the parameters of the inner disk (Table~\ref{tab_2}) are typical for a spiral galaxy, whereas the outer disk has a rather long radial scalelength and the $\mu_0$ that
is typical for low surface brightness (LSB) galaxies. In Mrk334 we caught an LSB disk in the
process of formation.  Here the mean brightness profile has already become close to an exponential, despite the asymmetry of azimuthal light distribution that is still very inhomogeneous.
The brightness of the outer disk should become more homogeneous after a few revolutions that is
after about $0.5-1$ Gyr.

Figure~\ref{f03}b shows the $R$-band brightness distribution after subtracting the model consisting of two disks and a bulge. A complex system of bright loops becomes immediately visible. Three characteristic radial scales can be identified: the circumnuclear ring with
a radius of about ($4-5$~kpc); inner filaments at a distance of $r=9-13$~kpc, and outer features  --- a loop and an arc located northwest and southwest of the nucleus, respectively, which can be observed beyond $r\sim 40$~kpc.  This subdivision of tidal features is rather arbitrary, we are most probably observing a disruption of a single galaxy torn apart by tidal forces and spread along its orbit around  Mrk 334. The bulk of the  companion stars is concentrated in the central region, because the mean brightness of the inner and circumnuclear filaments is twice higher than  that of the outer ones.

\begin{figure*}
\centerline{
\includegraphics[width=8cm]{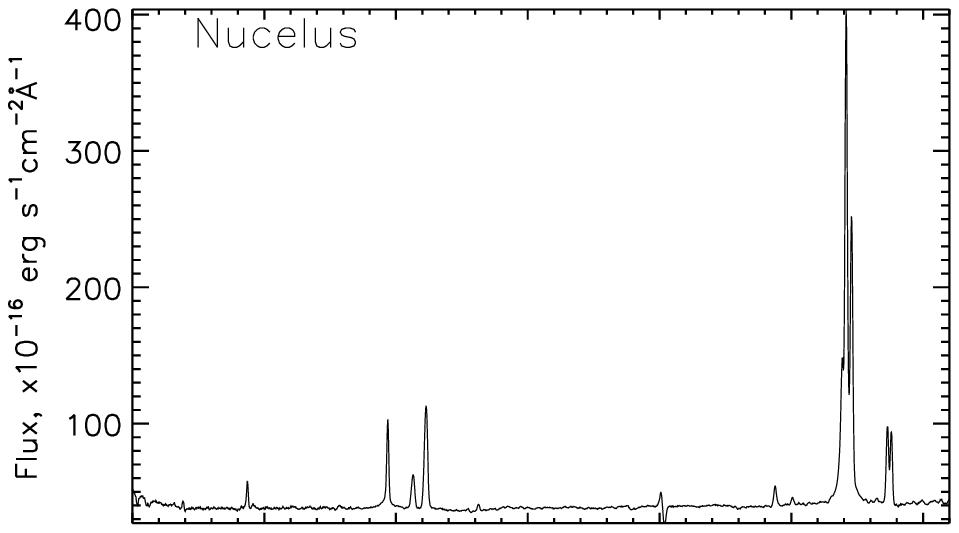}
\includegraphics[width=8cm]{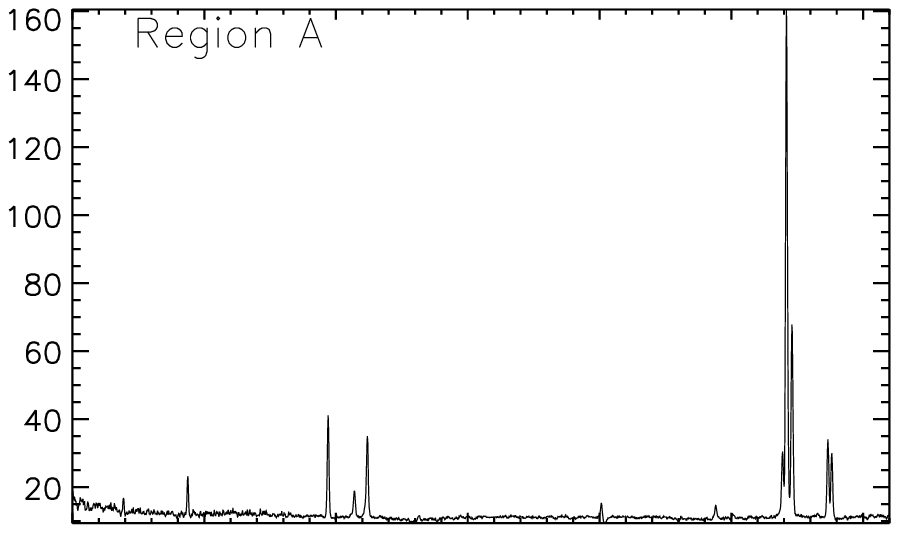}}
\centerline{\includegraphics[width=8cm]{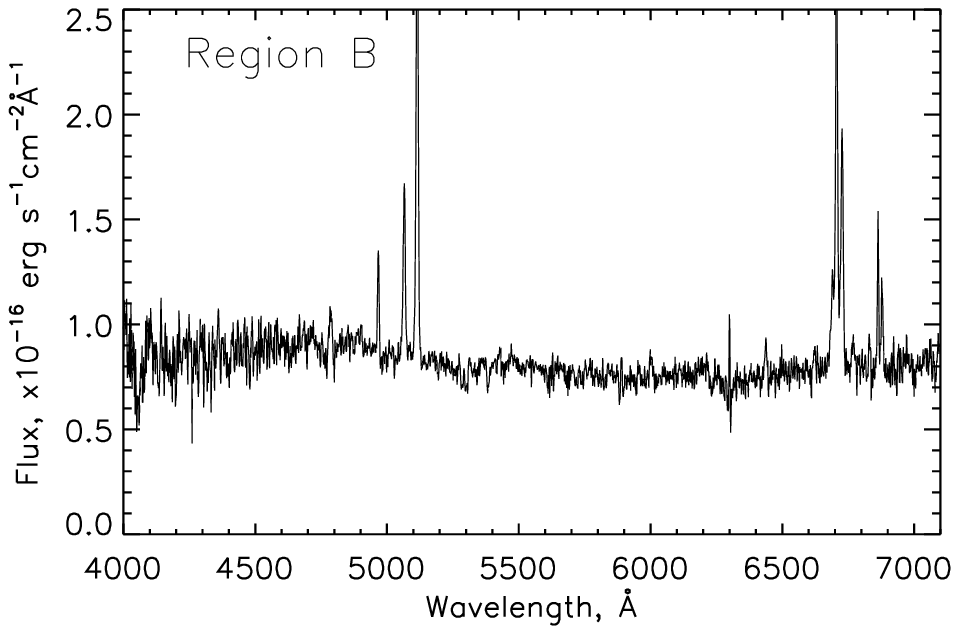}
\includegraphics[width=8cm]{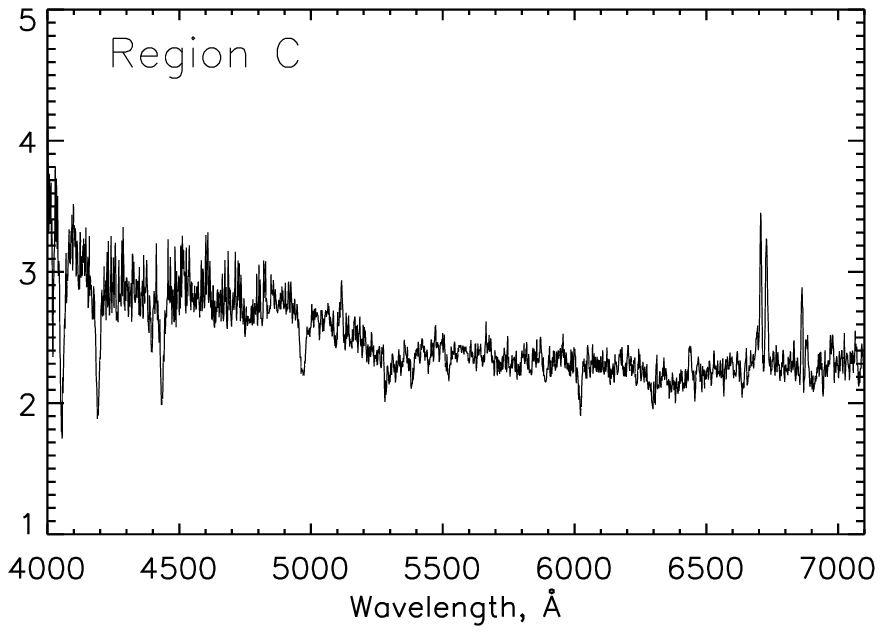}}
\caption{Integrated spectra of individual regions in Mrk 334.
}\label{f06}
\end{figure*}

The total luminosity contribution of all inner and outer filaments determined after subtracting the two-dimensional model is $30\%$ in the $V$  and $25\%$ in the $R$ band. Under the assumption that all filaments (including  the circumnuclear ones) were formed by stars of the satellite, we have derived the mass ratio for the galaxies  before interaction  ranging from   $1/5$ to $1/3$ for the equal $M/L$ ratios in both galaxies. The scatter of the estimates is mostly due to the uncertain fraction of companion stars dropped into the  outer disk.  This ratio is close to the conventional boundary between minor and major merging. Whereas in the former case the interaction should rather regarded as a simple accretion of a low-mass companion by the primary galaxy, in the case of major merging this process also distorts appreciably the structure of the more massive companion.

Figure~\ref{f05} shows the central part of  Mrk 334. The system of arcs and loops corresponding to the orbit of the disrupted companion shows up conspicuously on the map of residual brightness. According to the $(V-R)$  maps, individual fragments of these  loops  stand out because of their bluer colours, which are indicative of the presence of younger population that may have formed in the process of interaction between the galaxies. The red colour of Region `A' is evidently due to the H$\alpha$ emission  falls within the $R$-filter passband. Note that the tidal structures at $r=5-10$ arcsec are most likely located outside the plane of the galaxy, as follows from the analysis of the gas velocity field (Section~\ref{sec4}).

At the same time, the circumnuclear spiral at $r<5$ arcsec, which shows up on HST images  (Fig.~\ref{f05}c)  seems to belong to the galaxy disk plane. The authors  having reported HST images of Mrk 334 \citep{MartiniPogge1999,Martini2001} have also pointed out the minispiral and chaotic dust features in the circumnuclear region.

\section{Sources of Gas Ionization.}
\label{sec3}

Figure~\ref{f06} shows the integrated spectra of Regions `A', `B', and `C' and of the nucleus extracted from the MPFS data cube. The boundaries of these regions are shown in Figure~\ref{f05}.  The spectra of the nucleus and of the knot `A' have much in common: both exhibit strong Balmer lines and weaker
forbidden lines, mostly  [OIII]$\lambda\lambda4959,5007$\, and [OII]$\lambda3727$. Region `B' shows
the opposite pattern: the most conspicuous line is  [OIII]$\lambda5007$, which is even brighter than  H$\alpha$. The spectrum of Region `C' exhibits, along with a very weak H$\alpha$ emission, high-contrast MgI, Fe, and Ca  and Balmer-line absorptions. Such a spectrum is typical for a composite post-starburst region. We estimate the luminous-weighted age of the stellar population as 1.1-1.6~Gyr using the ULYSS\footnote{ULYSS is an open code located at \textrm{http://ulyss.univ-lyon1.fr/} and based on the papers of \citet{Koleva2009}, \citet{Chil2007} and other.} program package.

\begin{figure*}
\centerline{
\includegraphics[height=7cm]{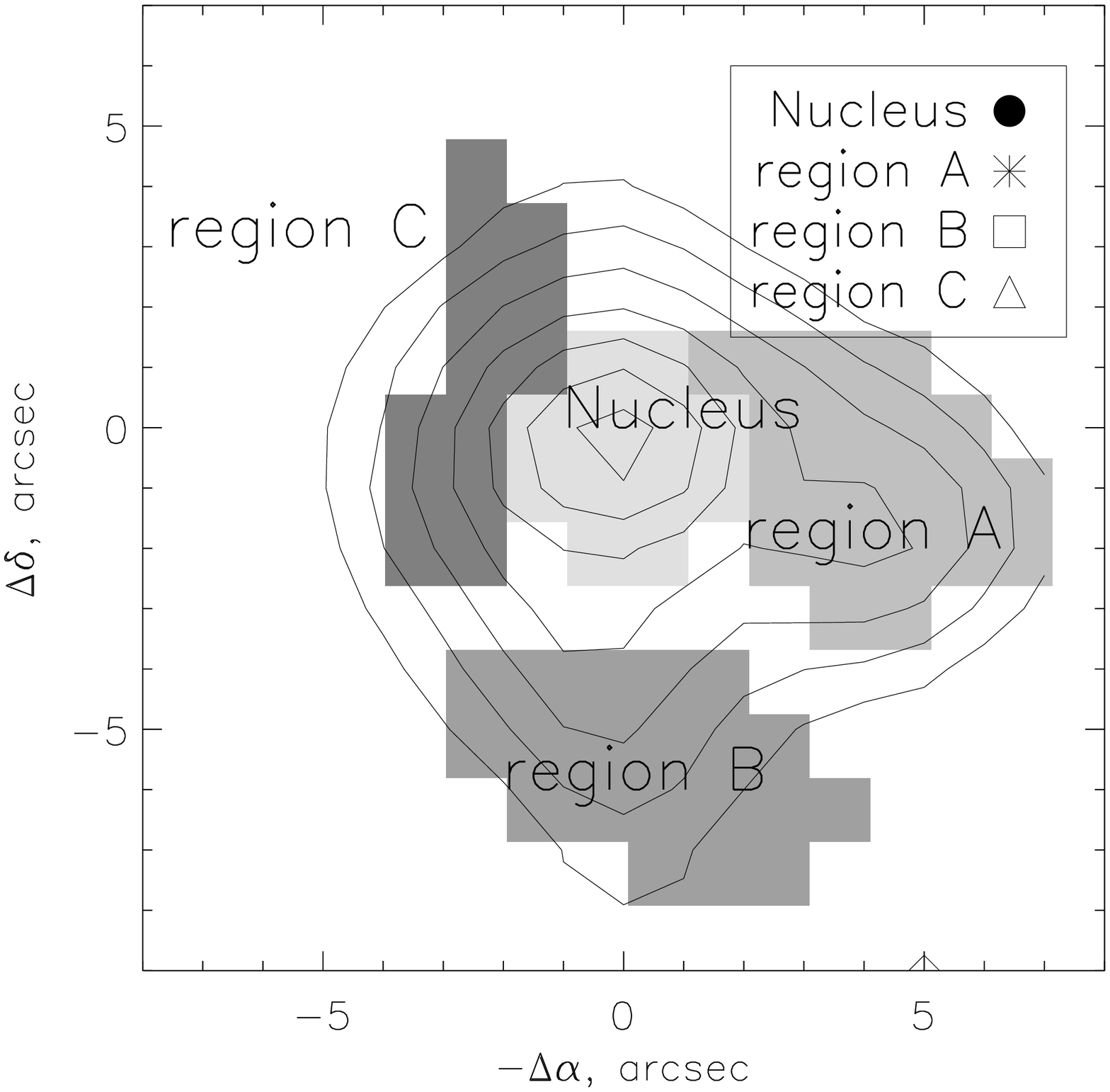}
\includegraphics[height=7cm]{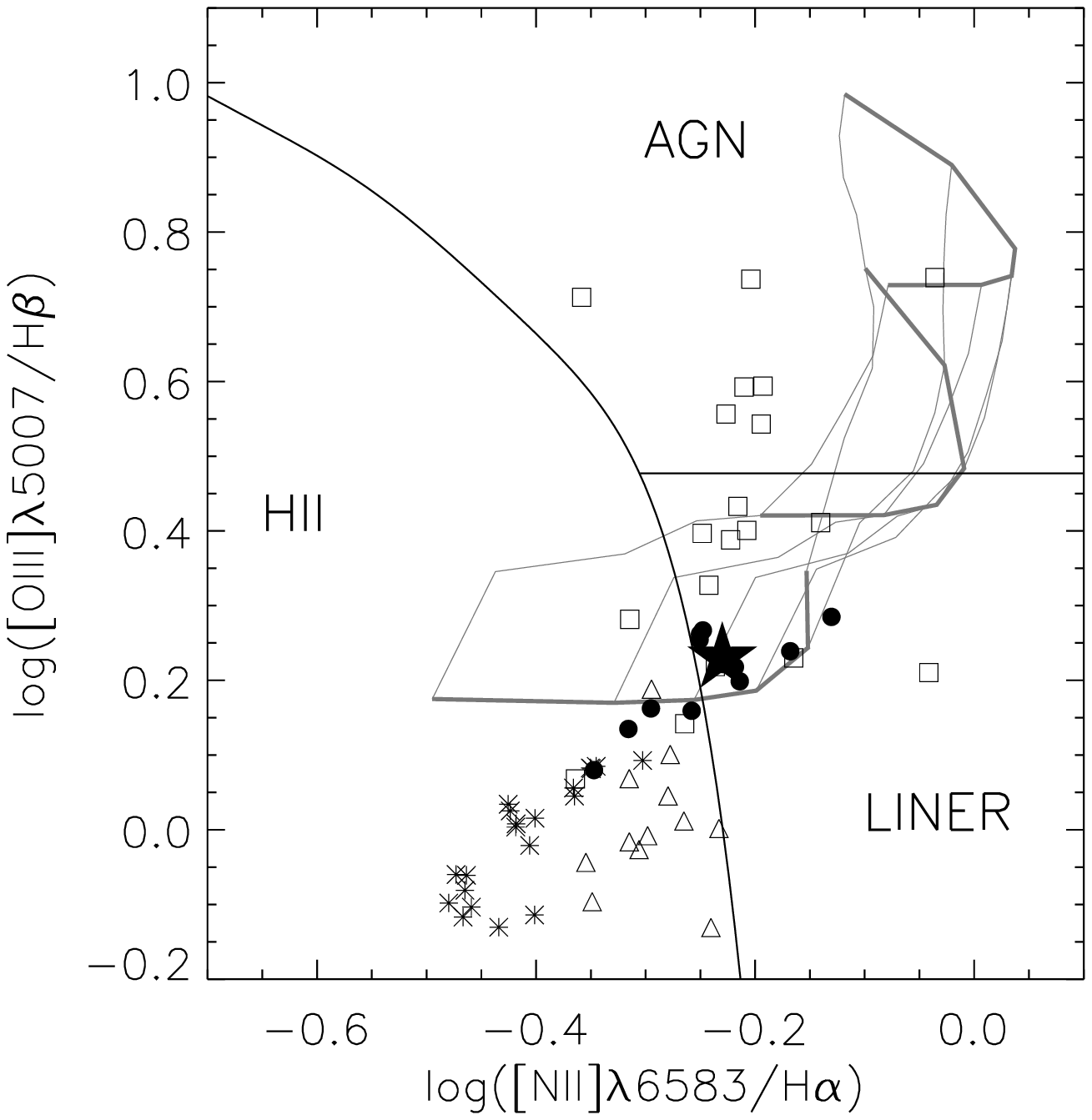}}
\centerline{
\includegraphics[height=7cm]{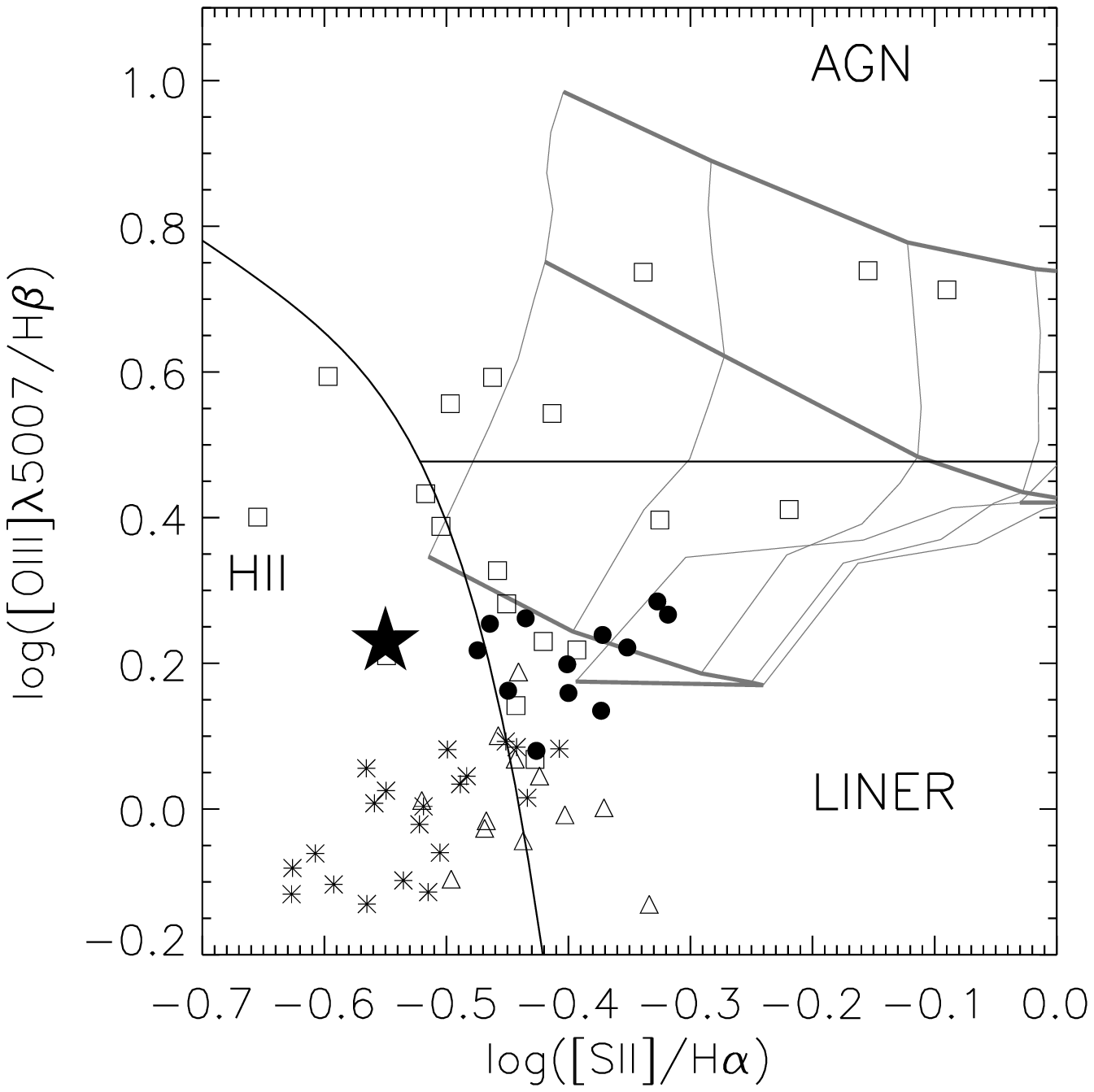}
\includegraphics[height=7cm]{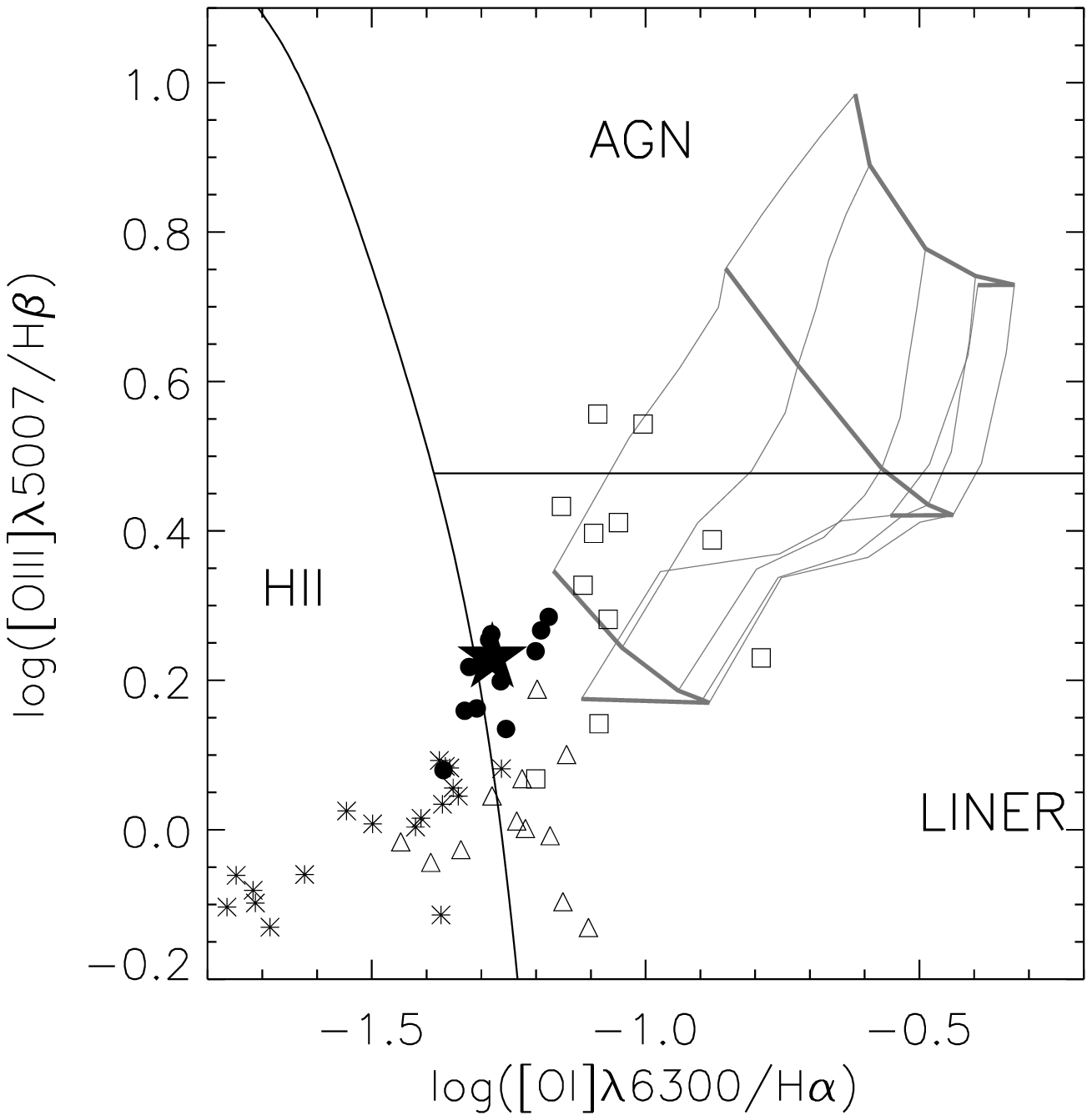}}
\caption{The diagnostic diagrams for various regions in  Mrk 334. The top left panel shows the mask used to select the galaxy regions,  the [OIII]-line isophotes are overlapped.  The circles, asterisks, squares, and triangles correspond to the nucleus and Regions `A', `B', and `C', respectively. In the case of Region `C' only  spaxel contained emission line spectra are shown. The black star shows the
data by \citet{Lumsden2001}.  The solid black lines separate the domains corresponding to different excitation mechanisms.  The gray lines show the grid of  shock+precursor models from \citet{Allen2008} for $n=1\,\mbox{cm}^{-3}$ and solar elemental abundances. The thin and bold gray lines mark the contours of constant magnetic parameter $B/n^{1/2}=0.1,0.5,1.0,2.0,5.0,10\,\mu \mbox{G}\mbox{cm}^{3/2}$, and
contours of constant velocity, $v=250,350, 450\km$ (velocities increase from bottom to top), respectively.
}\label{f07}
\end{figure*}

Regions with different emission-line spectra must also differ in their ionization sources. We construct the diagnostic diagrams  to determine the ionization mechanism for the inner  regions of the galaxy. Given the line ratios for
different excitation mechanisms, we can identify regions dominated by thermal (young  stars), nonthermal
(active nucleus), or shock ionization (hereinafter referred to as  HII, AGN,  and LINER). Figure~\ref{f07} shows the most typical diagrams. In the diagrams we adopted the boundaries separating domains corresponding to the different excitation mechanisms from \citet{VeilleuxOsterbrock1987}.
We ignored the effect of internal extinction, because  we use the intensity ratios of the lines with close wavelengths.

Mrk 334 is classified as a Sy 1.8-type galaxy and its Balmer-line profiles have a  low-contrast broad
component. We decomposed the H$\alpha$ and H$\beta$ line profiles into two Gaussian functions: one for the
broad and one for the narrow component. Only the  narrow component flux was used  in  the diagnostic diagrams.

The points in the diagrams of Fig.~\ref{f07} show the line ratios in each MPFS spaxel. In all diagrams the points corresponding to the nucleus lie at the  HII-LINER boundary.
Here the main ionization mechanisms are radiation of young stars and shocks, and not the nonthermal UV
continuum as it is typical of an active nucleus. The points corresponding to the nucleus are
located in the diagrams  so far from the  LINER-AGN boundary  that the nucleus of  Mrk 334 should be classified as a
LINER rather than a Sy galaxy. This conclusion is consistent with the recent spectrophotometric studies of the nucleus reported by  \citet{Lumsden2001}. The asterisks in Fig.~\ref{f07} indicate the line ratios in the nucleus as inferred from their long-slit spectrum\footnote{The fluxes reported by \citet{Lumsden2001} are
extinction corrected, but, as we point out above, this correction is negligible for the diagrams}.
Figure~\ref{f07} demonstrates the good agreement between their data and our measurements for the nucleus except for the  [SII]/H$\alpha$ ratio. The last discrepancy must be due to the fact that we corrected our spectra for the atmospheric absorption band which partially overlaps with the [SII] lines and decreases sulfur flux by $\sim0.1\,\mbox{dex}$.  \citet{Lumsden2001} say nothing about applying such a correction.

The emission-line ratios observed in the nucleus can be explained in terms of the following supposition.
The star formation in  the nucleus   is so violent that the total line
emission is determined mostly by the collective effect of photoionizing radiation of young stars and
by the shocks produced by supernova explosions. At the same time, the emission lines  of the active nucleus
are barely visible against the circumnuclear starburst. The weak broad component with $FWHM\approx2500\km$
in the hydrogen line profiles and FeII features in the spectrum of the nucleus are the only indications of
an AGN central engine.

In all diagrams the points belonging to Knot `A' lie deeply in the region corresponding to the ionization by  OB stars  radiation. Thus  `A' is indeed a region of intense ongoing star formation. It accounts for  $15-20\%$ of the total  H$\alpha$ luminosity, and such a fraction
formally corresponds to a star formation rate of  $SFR=3\,\mbox{M}_\odot/$yr. Such
a high value (equivalent to the total SFR in the starburst galaxy M~82) in a relatively compact region (1 kpc in  size) is indicative of  a powerful starburst.

\begin{figure}
\includegraphics[height=4.2cm]{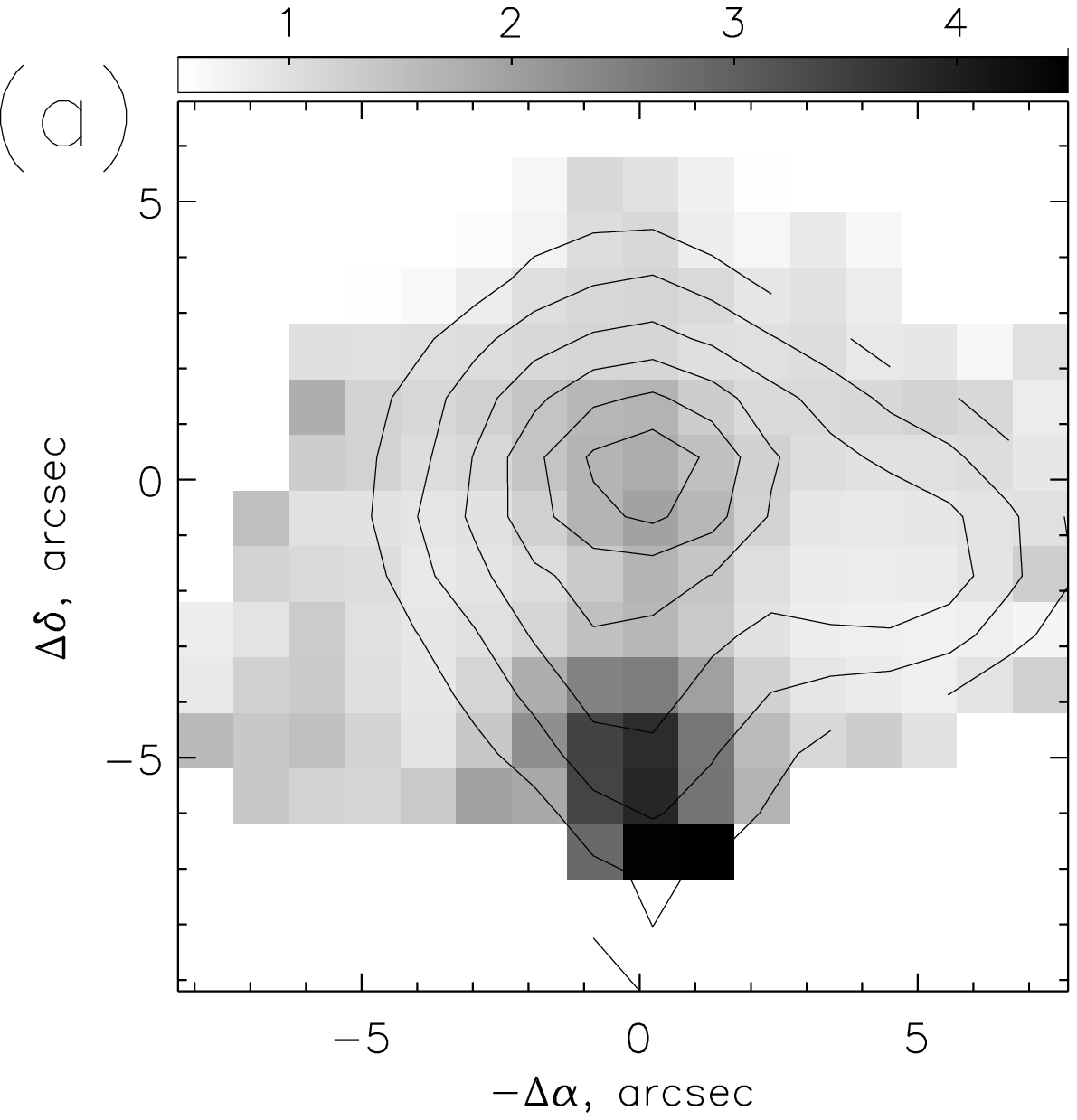}
\includegraphics[height=4.2cm]{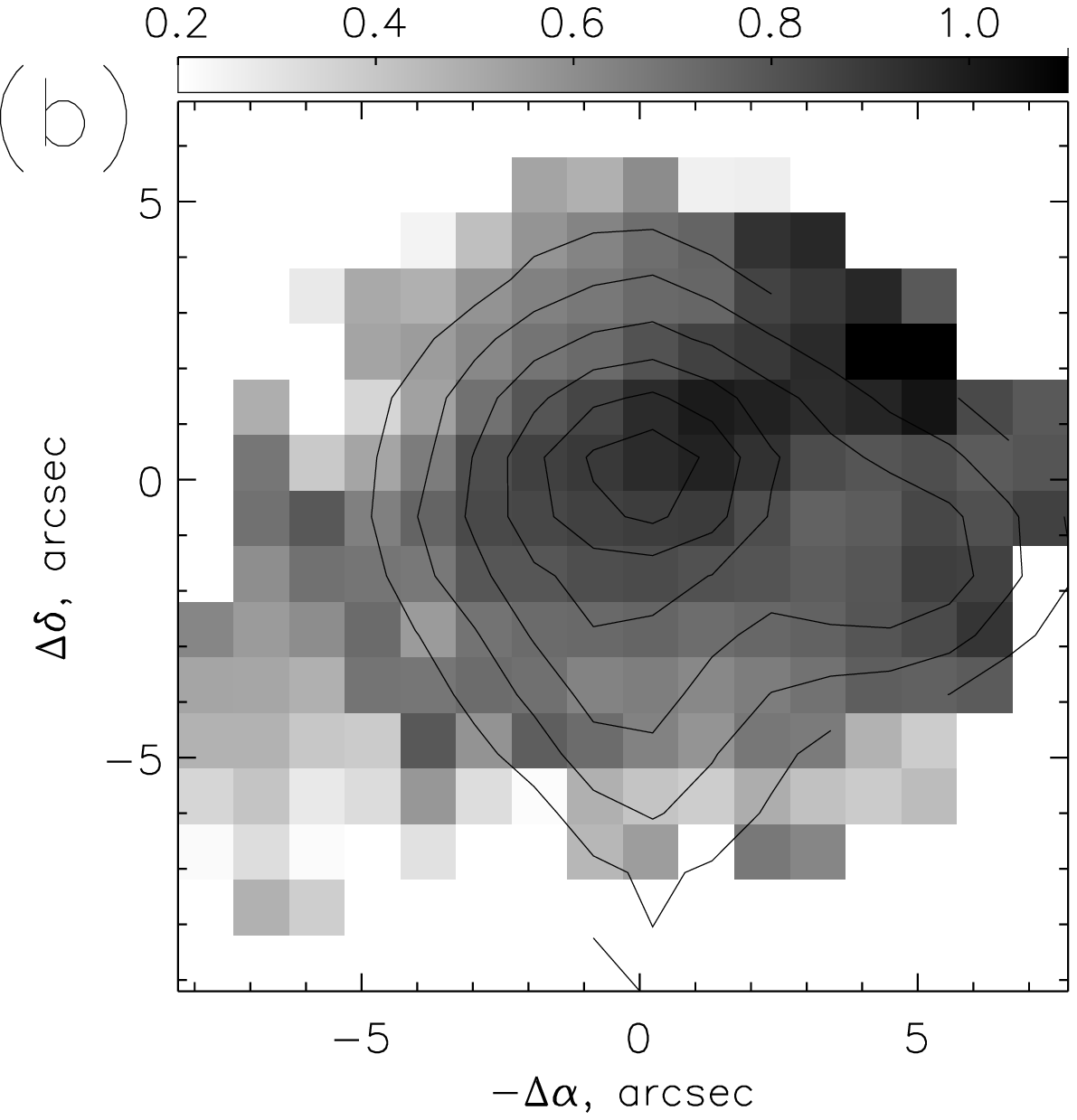}
\caption{Map of the  [OIII]$\lambda5007$/H$\beta$ (a) and
[SII]$\lambda6731/6717$ (b) line ratios. The [OIII]$\lambda5007$ isophotes are overlapped. In the case of sulphur lines, darker colours
correspond to higher electron density.}\label{f08}
\end{figure}

Condensation `B' appears to be the most intriguing among these features:
it exhibits unusually high [OIII]$/\mbox{H}\beta$ line intensity ratios (see Fig.~\ref{f08}a)that formally correspond to the ionization by an AGN. Two hypotheses can be suggested to explain this peculiarity. First, Region `B' may be the active nucleus of the satellite. \citet{Barth2008} recently observed a similar situation in the interacting galaxy
NGC\,3341, where the spectrum of the nucleus of the disrupted companion exhibits
Seyfert-type features. However, Condensation `B' is barely visible on
the optical continuum images of  Mrk 334. Observations made at the 6~cm wavelength  \citep{Ulvestad1986} with an angular resolution of  $0.4-0.6$ arcsec also demonstrate the lack  of a nonthermal radio emission typical of an active nuclei in Region `B'.

The second hypothesis is based on the fact that Region `B' is located
close to the tidal arclike structures identified in the circumnuclear region images (Fig.~\ref{f05}b). It would be safe to assume that here we see the intersection between the galaxy disk and the orbit of the disrupted companion  remnants. It is the locus where
the debris have `punched' the gaseous disk of Mrk 334, thereby
creating in Region  `B' a cavern with lower than ambient gas density.
The high degree of ionization of the gas is due to a powerful shock. This
idea is also supported by the electron-density estimates
derived from the  [SII]$\lambda\lambda6730/6717$ line flux ratio using the
relation adopted from  \citet{Osterbrock1989} for $T_e=10\,000\,K$. The electron
density is equal to  $n_e=250-430\,\mbox{cm}^{-3}$ in the nucleus,
$n_e=200-350\,\mbox{cm}^{-3}$ in Region `A', and increases to
$500\,\mbox{cm}^{-3}$ to the north of this  HII region (Fig.~\ref{f08}b). At the same
time, the sulphur line ratio is less than 0.7 at almost all points of
Condensation `B', which is impossible to interpret in terms of simple
photoionization models. It is indicative of high temperature and low gas
density ($n_e<20\,\mbox{cm}^{-3}$).

We tried to compare the line ratios observed in Region `B' with the
results by the modern shock ionization simulations adopted from \citet{Allen2008}. We found that the theoretical shock+precursor model
predictions   describe fairly well the line ratios observed in Region `B' (Fig.~\ref{f07}) for parameters  that   can be reasonably expected for the interstellar medium:  the density of $1\,\mbox{cm}^{-3}$, solar elemental abundances and magnetic-parameter
$B/n^{1/2}=0.1-10\,\mu \mbox{G}\mbox{cm}^{3/2}$. The required shock speed is $250-350\km$.
The shock origin of the [OIII] emission  is also evident from the line broadening. Indeed the width of the [OIII] lines in Region `B', after instrumental width correction, corresponds to the  $\sigma=160-200\km$ instead $60-100\km$ in other disk points, with the exception of the nucleus. More kinematic evidences will be presented in the next Section.

Region `C' may be satellite debris that has recently punched a hole in the gaseous
disk of  Mrk 334. It is located close to Region `B';  it appears sufficiently bright in the continuum
images and its spectrum contains   lines of young stellar population; it appears compact compared to other nearby filaments. It may actually be the remnant of the nucleus of a companion galaxy. The data on the ionization sources in this region are rather scarce and available only for the boundary between Region  `C' and the nucleus of Mrk 334 because the spectrum shows mostly stellar absorptions. In the diagnostic diagrams Region `C' lies in the  HII and LINER sectors. The photoionization  is caused here mostly by star-forming processes, and the shock ionization contributes appreciably in the points located close to the nucleus.

\section{Kinematics of Ionized Gas and Stars.}
\label{sec4}

\begin{figure}
\includegraphics[width=8cm]{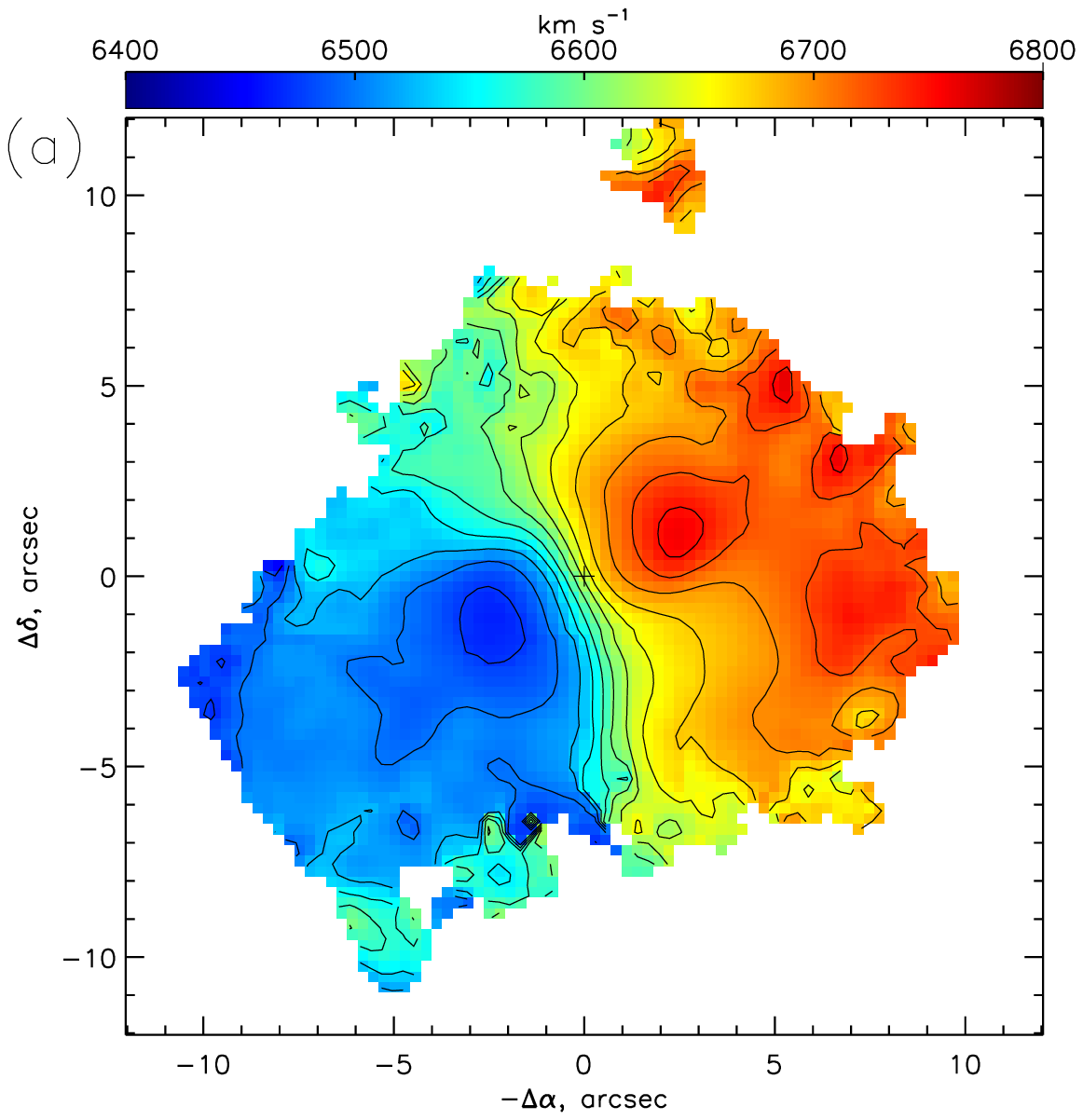}
\includegraphics[width=8cm]{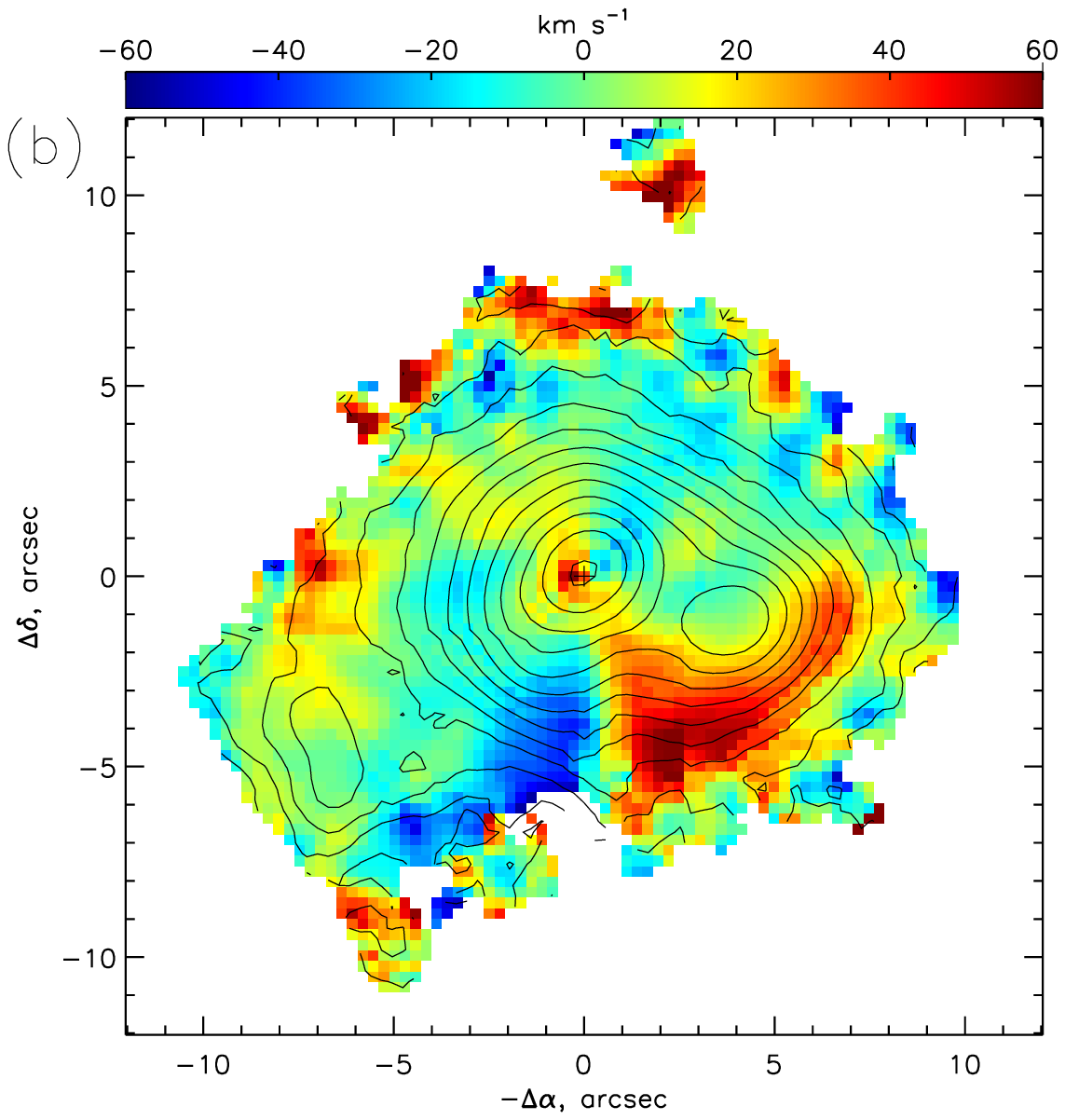}
 \caption{Kinematics of ionized gas according to the H$\alpha$-line FPI data:
the  velocity field (a) and   the residual velocities (observations minus model) with H$\alpha$-line
isophotes overlapped (b).}\label{f09}
\end{figure}

\begin{figure}
\includegraphics[width=8cm]{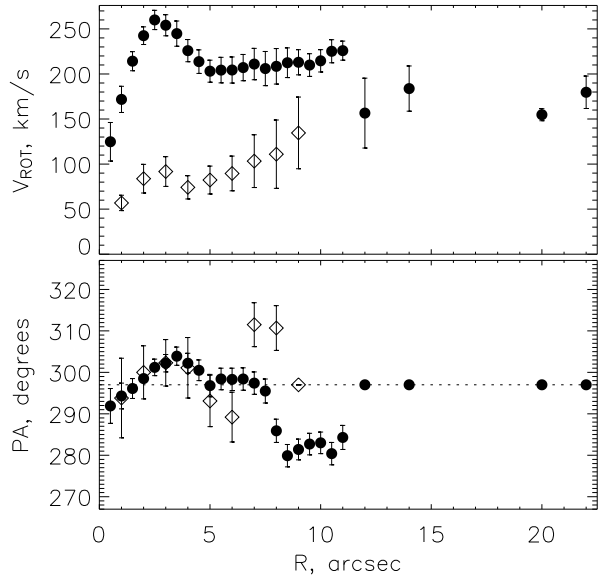}
\caption{The rotation curve (top) and variations of the position angle of the
major axis with a radius (bottom). The black circles show the
H$\alpha$-line FPI velocity-field measurements and diamond signs mark the
measurements for  the stellar component (MPFS). }\label{f10}
\end{figure}

The H$\alpha$  velocity field derived from the FPI data seems to be in a good agreement with the model of a regular rotating thin disk (Fig.\ref{f09}). We fitted the velocity field by the ``tilted-ring'' model  using the   algorithms  employed earlier to study NGC 6104 \citep{Smirnova2006}. The circular rotation explains the gas velocity field fairly well.  Therefore we
think that the inner tidal features ($r=5-10$ arcsec)  are located outside the galaxy plane and do not perturb the entire gaseous disk but only cross it in some places.  We found the disk inclination to be $i_0=34pm6$~deg and the line-of-nodes  position angle of $PA_0=297\pm3$~deg. Fig.~\ref{f10} shows the rotation curve of ionized gas ($V_{rot}$) and the radial variations of the kinematic axis ($PA_{kin}$). The data points in the range of $r=12-22$ arcsec come from three outer HII regions located far from the central disk:  $\mbox{PA}_{kin}$ could not be determined from these regions, and we assume it to be equal to the mean $PA_0$ of the disk.

The same figure shows kinematic parameters for the old stellar population. The rotation velocity measured for stars is about twice  (by $\sim100\km$) smaller than that found for
the ionized gas. This discrepancy
must be due to asymmetric drift, because  the central velocity dispersion of stars reaches, according to our estimates, $170-200\km$. The rotation curve of both  gas and stars exhibits a characteristic peak near the effective radius of the bulge.  At greater galactocentric distances, $r=4-12$ arcsec, the rotation
velocity of gas is almost constant and equal to $210-220\km$. The $V_{rot}$ of the external HII regions mentioned above is lower by  $40-50\km$, however, we have no grounds to believe that these regions may be located outside the disk of the galaxy. It seems to be more likely that  the formal decrease of the rotation velocity results from  a  contribution of non-circular gas motions.

Small variations of $PA_{kin}$  at $r<5$ arcsec indicate  the influence  of the circumnuclear spiral on the kinematics of the gaseous and stellar subsystems. The  $PA_{kin}$ abruptly deviates from the line of nodes at $r=8-11$ arcsec. Such a behaviour is indicative of large-scale noncircular motions at the edge of the HII disk. The  residual-velocity map (Fig.\ref{f09}b)  shows the distribution of observed velocities  after the subtraction of the model. Deviations from circular  rotation are small ($15-20\km$) in the regions with the brightest H$\alpha$ emission. However, to the south from the
nucleus an extended region can be seen where peculiar velocities are much higher and vary smoothly from
$-70$ to $+60\km$  in the east--west direction. Region `B' is located here, which we have identified
earlier by its spectrophotometric properties, primarily by its high [OIII]/$H_\beta$ ratio. Now we see that this region is also distinguished by the peculiar kinematics of ionized gas. Such a velocity distribution corroborates the hypothesis --- suggested above -- that Region `B' is the locus where the debris of the disrupted companion crossed the gaseous disk of the galaxy.

An alternative explanation of the observed gas kinematics is a jet--clouds interaction similar to that observed in  Mrk~3 \citep{Capetti1999} or Mrk 533 \citep{Smirnova2007}. However, this mechanism is unlikely for  Mrk 334. Firstly, as we already pointed out above, this galaxy has no extended radio structure. Secondly,
if the jet acts on the interstellar medium then the expected gradient of peculiar velocities should be directed away from the nucleus, i.e., in the radial and not in the azimuthal direction as we see in Fig.\ref{f09}b.

\begin{figure}
\centerline{\includegraphics[height=4cm]{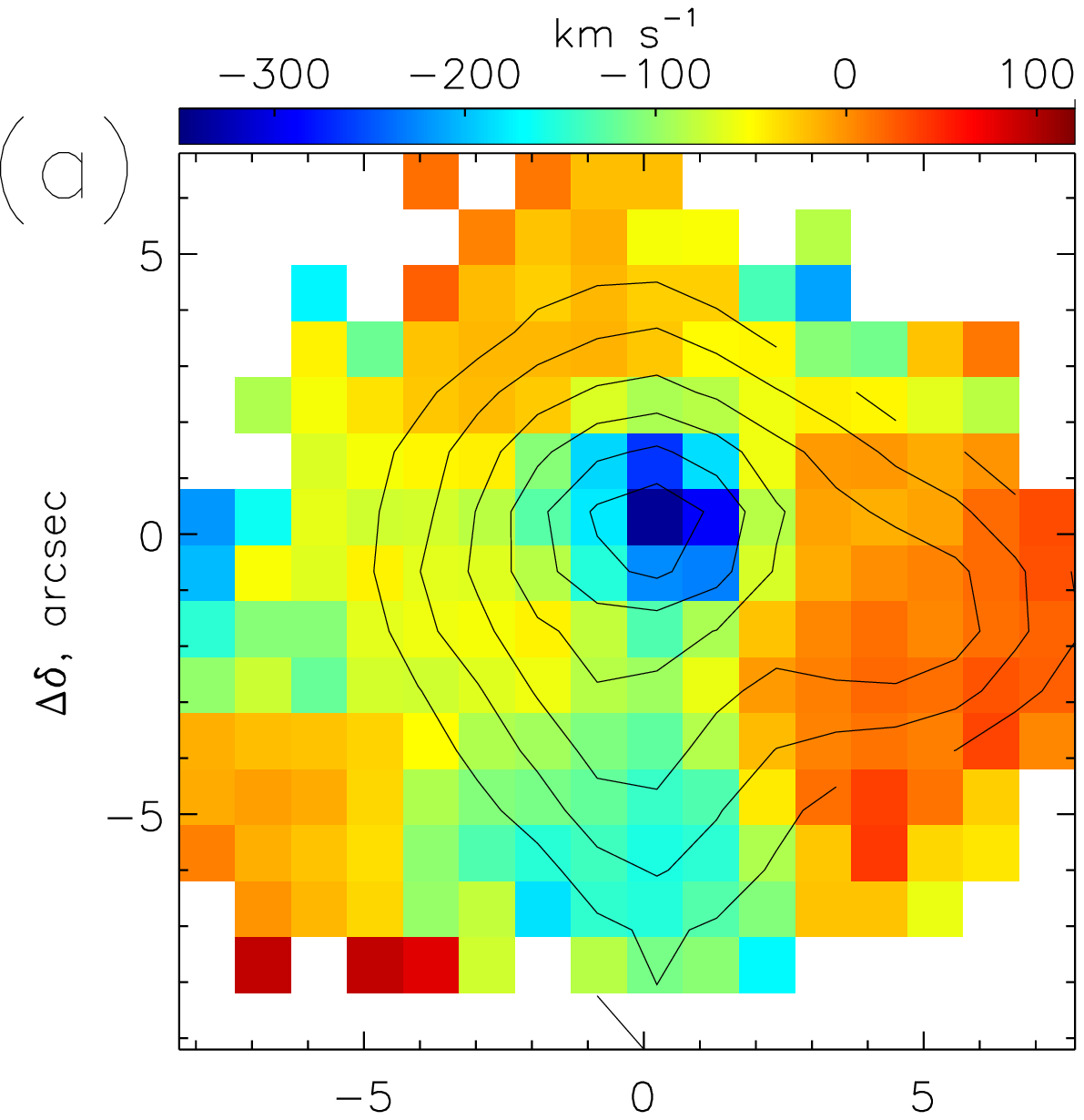}
\includegraphics[height=4cm]{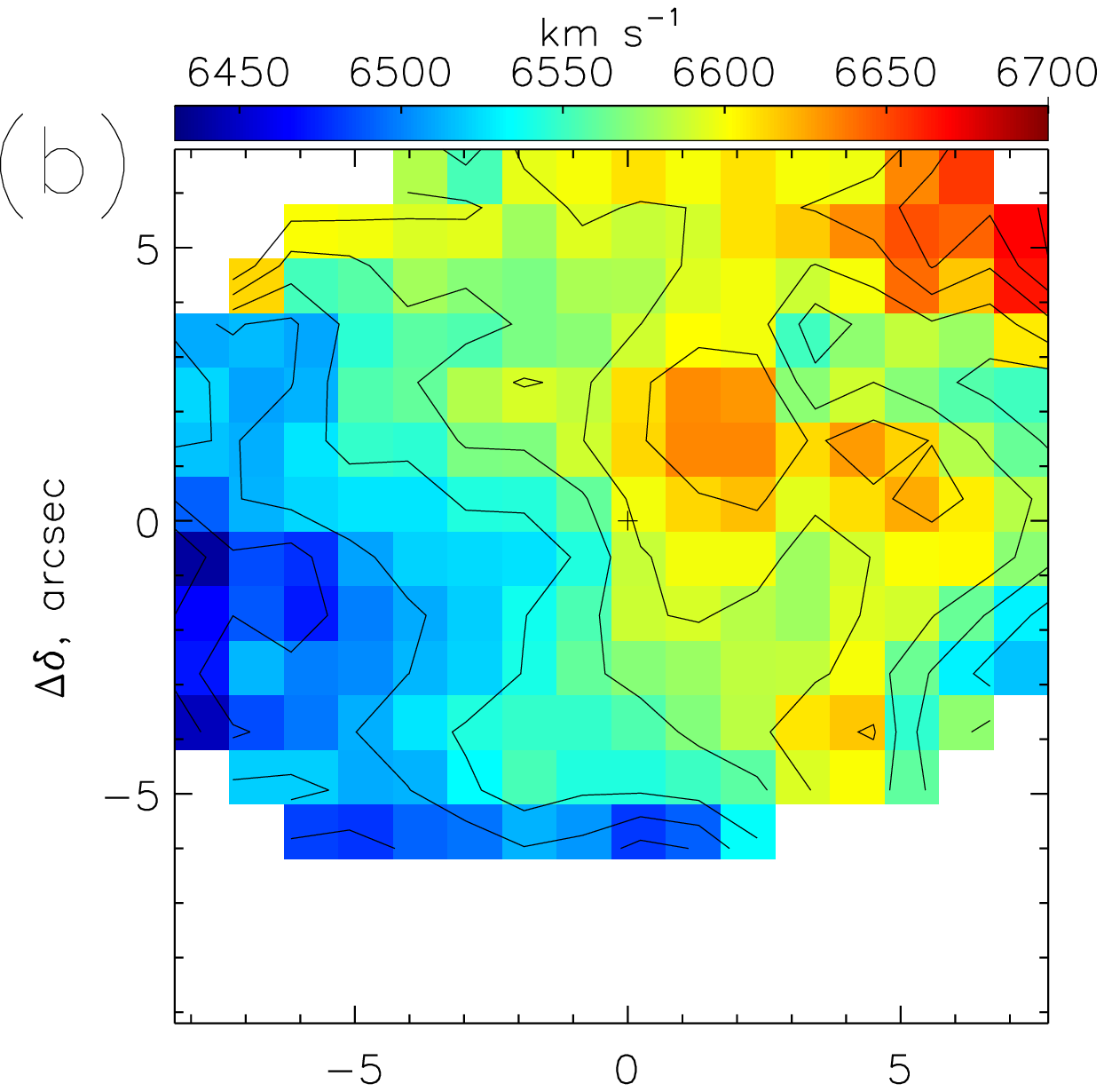}}
\centerline{\includegraphics[height=4cm]{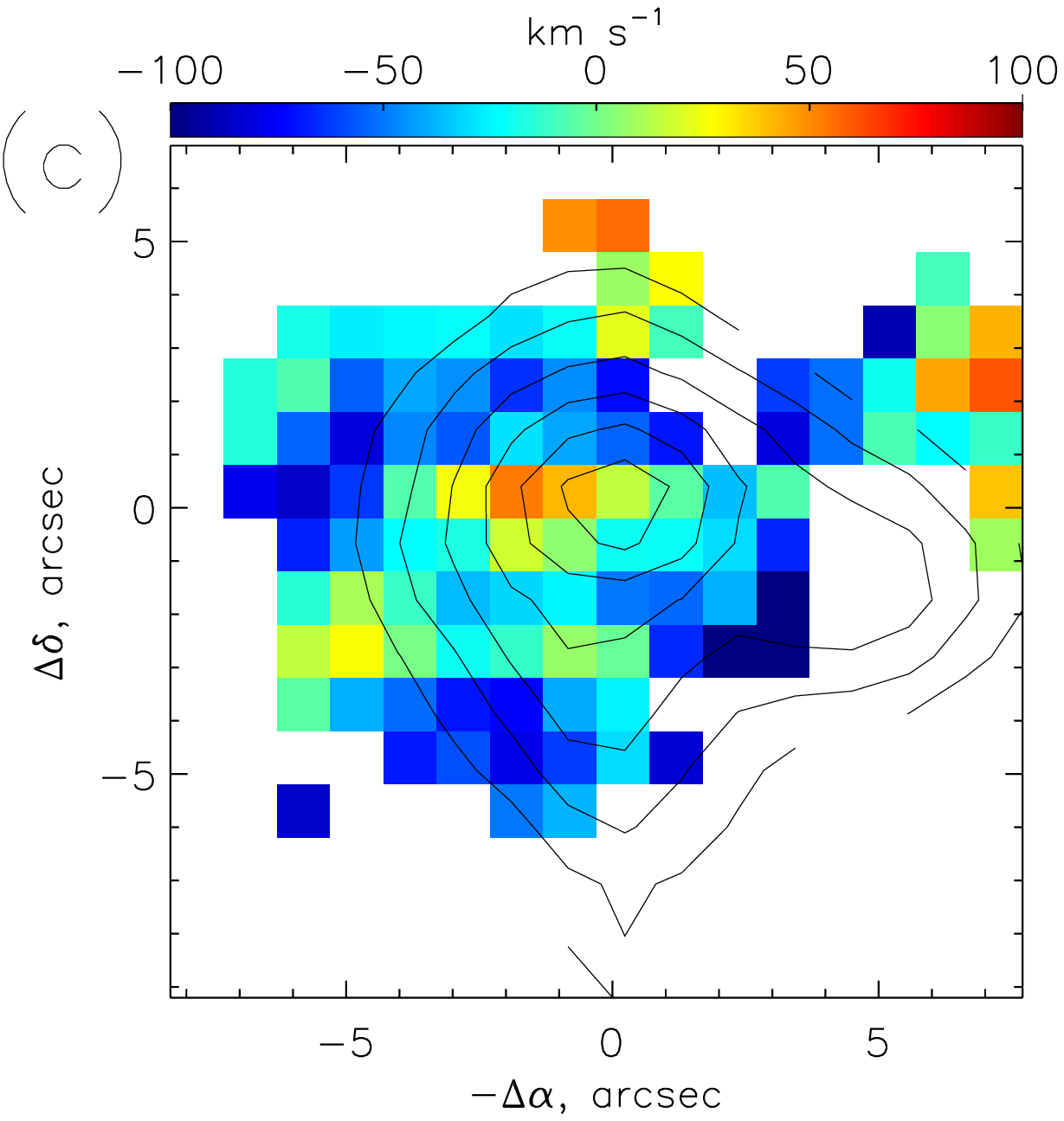}
\includegraphics[height=4cm]{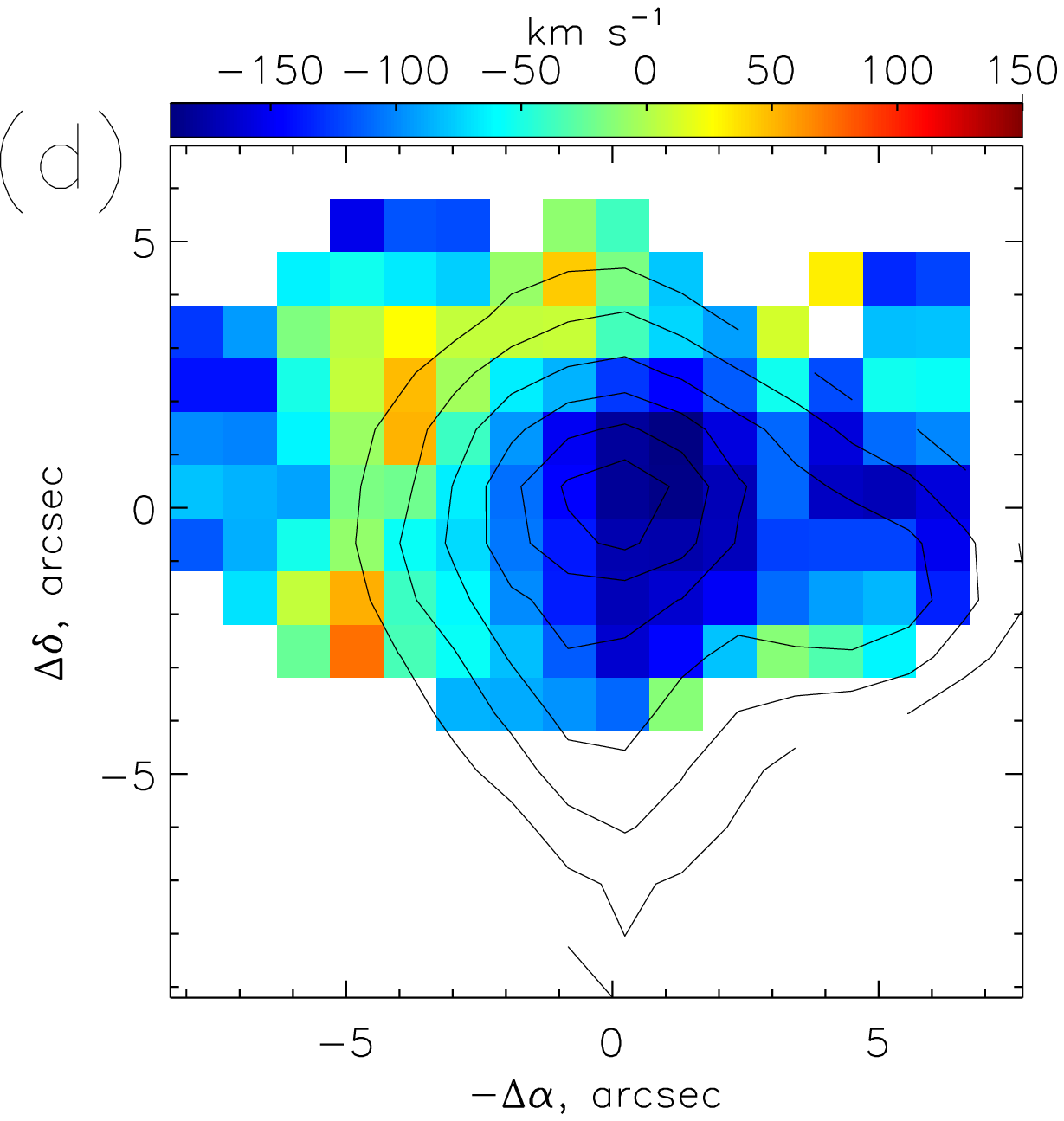}}
\caption{(a) Residual  [OIII]-line velocities (observed velocities
minus the rotation model in H$_\alpha$). (b) -- Velocity field of the old
stellar population. (c) -- The difference between the velocities of the
`young'  (F-type stars) and `old' (K-type stars) populations. (d) -- The
residual velocities in the NaD line minus the rotation velocity of the old
stellar population. Figures~(a), (c), and (d) are shown with the [OIII]-line
isophotes superimposed.} \label{f11}
\end{figure}

The velocity fields derived from the MPFS data  (Fig.\ref{f01}) have allowed us to study the kinematics of gas in lines excited by different mechanisms, albeit with lower accuracy and in coarser
detail compared to the results based on the FPI H$\alpha$ data. The gas motions
observed via most of the low-excitation emission lines ([OI], [OII], [NII], [SII]) agree well with the
picture found in the  $H_\alpha$ line. Namely, they  show a circular rotation with appreciable deviations near Region `B'.
Only the velocity distribution in the [OIII] line differs from the overall pattern (Fig.\ref{f01}, bottom). Figure~\ref{f11}
shows the residual velocities in this line  after the subtraction of the circular-rotation model derived from the   H$\alpha$ data. Two features are apparent. First, the residual velocities in Region `B' reach $-150\km$, which is  greater by amplitude than the corresponding velocities for the low-excitation lines. Second, the galaxy nucleus shows a significant excess of negative velocities (down to $-300\km$ ).  Similar gas outflows from  AGN   (excess of `blue' velocities, first and foremost, in the  [OIII] line) were found in the integral-field spectroscopy data for Seyfert galaxies studied by us and by other teams: Mrk 315 \citep{Ciroi2005}, NGC 2273 \citep{Moiseev2004}, NGC 2992 \citep*{Garcia2001},  NGC 6104 \citep{Smirnova2006} and others. A nuclear-blueshifted outflow associated with this feature is usually interpreted as a jet--clouds interaction \citep{Ferruit2002}, or, in a more general case, as  hot wind emerging from an active nucleus (\citet{Komossa2008} and references therein). However, the
situation in Mrk 334 must be different. Firstly, no radio jet can be seen in high-resolution radio maps,
in any case, its size cannot exceed $0.5$  arcsec. Secondly, in the nucleus itself the contribution from the
nonthermal component to the ionization of gas is smaller than that of star formation (see Section~\ref{sec3}). Thirdly, unlike the examples of active galaxies mentioned above,  the
high-velocity outflow  in Mrk\,334 is observed only in the  [OIII] line. This outflow is most likely associated with intense star formation in the nucleus rather than with the central engine itself. Thus, what we observe in the [OIII] line is a low-density gas ejected above the plane of the galaxy as a result of multiple supernova explosions -- the so-called `superwind'. Below we analyse this possibility in  more detail.

We computed the velocity field of the stellar component by correlating the
spectra of the galaxy with the stellar spectra from the MILES library and
selecting the spectral type of the template and wavelength interval so as to
maximise the amplitude of the cross-correlation function. Figure~\ref{f11} shows the
velocity field corresponding to the old stellar population determined by cross-correlation
of the galactic spectra with the spectrum of a K0III type star in the interval of
$\lambda5120-5800$\AA.  This
velocity field was used  to construct the stellar rotation curve presented in Fig.~\ref{f10}. However, a significant contribution of younger stellar population can be seen in the bluer part of the galaxy spectra in  some regions. Thus in the
wavelength interval $\lambda3750-4350$\AA\, the coefficient of correlation
with the spectrum of a F--type star exceeds the corresponding value for a K--type star in the $\lambda5100-5500$\AA wavelength interval. The measurements made for the `old' and `young' populations yield
different line-of-sight velocities. Figure~\ref{f11} shows the difference
between the velocity fields of old and young stars. In the nucleus the difference is small
and does not exceed $20\km$, which is comparable to measurement errors. However, the velocity
difference reaches $-80\km$ in two regions to the east of the nucleus. Both regions
identified in the field of stellar residual velocities coincide with the
inner tidal loops in the circumnuclear region, including Region `C',  (Fig.\ref{f05}b).
These facts lead us to suggest
that here we see two kinematic components along the same line of sight. The old
stellar population belongs to the disk of  Mrk 334 and exhibits normal circular
rotation. At the same time, the tidal filaments formed in the process of
the companion disruption are dominated by younger population
(as a result of a relatively recent burst of star formation). The
filaments are located outside the disk plane as is evident from their
line-of-sight velocities. Fig.\ref{f11}c also shows another region with significant negative differences of the young and old stars velocities to the west from the nucleus. It may also include  stars from the companion galaxy.

An interesting pattern emerges if we cross-correlate the spectra in the NaD doublet spectral
domain. Figure~\ref{f11} shows the line-of-sight
velocities for this line measured after the subtraction of the velocity
field of the old stellar component. An excess of negative residual velocities up to $-180\km$ is immediately apparent in the nucleus of the galaxy. The NaD line is present not only in the spectra of late-type stars but also in the spectra of the interstellar medium. It is reasonable to associate the excess velocities in this line with the same superwind that we found in the [OIII]-line data for the ionized gas. We should note that outflow velocities may be underestimated because the contamination of the NaI line by absorption from the stellar population is also present.

\section{X-Ray Radiation of Mrk 334}
\label{sec5}

\begin{figure}
\includegraphics[width=8cm]{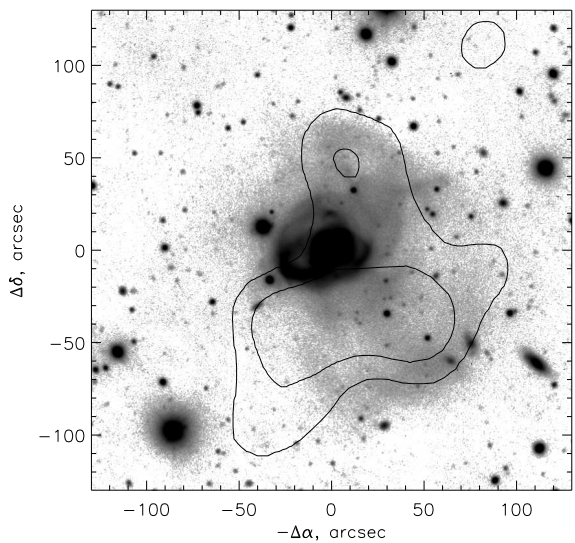}
\caption{The $R$-band image of the galaxy with \textit{ROSAT} X-ray isophotes from \citet{Zimmermann2001} superimposed.}
 \label{f12}
\end{figure}

Let us now briefly discuss the peculiarities of the distribution of X-ray flux of  Mrk 334 according to the
\textit{ROSAT} data.  \citet{Zimmermann2001} report isophotes of the smoothed image in the energy interval of $0.1-2.4$ keV.
The X-ray source with the luminosity of  $L_X=2.6\cdot10^{42}\, \mbox{erg}\,\mbox{s}^{-1}$ is unambiguously identified with the galaxy. However, the contours of the X-ray image are appreciably offset
with respect to the optical nucleus, and outer isophotes coincide with the tidal structures at $r=70-100$ arcsec northwest and southwest of the centre of the galaxy (Fig.~\ref{f12}). Hence  the diffuse X-ray emission is associated with the merging galaxy system but not with the Seyfert nucleus. Especially striking is the almost exact coincidence of the outer X-ray isophotes with the  edges of optical filaments which is surprising given the relatively low spatial resolution of the  \textit{ROSAT} data.

If the most of the X-ray radiation of  Mrk 334 is unassociated with the active nucleus, it may be due either to unresolved stellar sources or to the outer hot gas. We believe the former variant to be unlikely. Excess number of X-ray
point sources -- close binaries, ULXS, young supernova remnants associated
with a starburst -- are observed in a number of interacting galaxies. In this case other signs of ongoing star formation and, in particular, HII regions, should also be apparent at the periphery of the galaxy. However, according to our data,  H$\alpha$ emission is concentrated only in the central region, inside  $r<12$ arcsec.

If the source of the emission is hot gas then where does it come from? It cannot be the corona of the disrupted companion like those observed in giant  elliptical galaxies, because the mass of the companion is not too large, and Mrk 334 is a disk galaxy. An HI corona may have formed from the gas scattered
as a result of interaction, whereas most of this gas is concentrated in the disk plane. It is, however, unclear how this gas can be heated.

We could suggest only one more or less realistic scenario to explain  the formation of the extended X-ray structure. The inner parts of the galaxy are marked by high star-formation rate. Our estimates yield SFR=$18\,$M$_\odot\,\mbox{yr}^{-1}$ (from the H$\alpha$ line) and $12\,$M$_\odot\,\mbox{yr}^{-1}$ (as inferred from the \textit{IRAS} FIR  using the relations  from \citet{Kennicutt1998}). Such an intense star-forming activity in a compact region accompanied by supernova explosions may give rise to a superwind phenomenon that results manifests significant heating of the interstellar medium and its outflow within a wide cone perpendicular to the galaxy disk \citep*{Heckman1990,Veilleux2005}. Inside the cone the temperature may rise to $10^7$ K,
and hence the gas should be a powerful X-ray source. The starburst starts early enough during the interaction of galaxies  --- it begins before the complete disruption and merger of the companion. As an example, we can
mention the well-known superwind galaxy M~82 that is currently in the process of a tidal interaction with M~81, or
the NGC 6285/6286 galaxy pair \citep{Shalyapina2004}. Thus, a hot gaseous `bubble' or cone could be
formed above the plane of Mrk 334 even before the final stage of the merging. The spatial distribution of the hot gas
was then distorted because of the significant perturbations of the gravitational potential of the system. We
observe the result of these distortions as the offset of the X-ray brightness centre in the sky-plane projection.

We have already pointed out in Section~\ref{sec4} that the MPFS spectra are indicative of the presence of the superwind in Mrk 334. Gas motions directed toward the observer have been found in the central region.  The observed velocities ($180-300\km$) are typical of superwind galaxies with intense star formation in their nuclei \citep{Veilleux2005}.

\section{Discussion.}
\label{sec6}

The  analysis of various observational data lead us to conclude that  Mrk 334 is in the process of  merging with a companion that has already been almost completely disrupted by the tidal forces. Is the nuclear activity associated with such a close interaction? Let us first
turn to the galaxy morphology. The inner region  ($r\leq3$ kpc) hosts a well-defined
spiral pattern (Fig.~\ref{f05}c). Nuclear spirals in Mrk 334, where sites of star formation are located, are
relatively brighter than similar features in other galaxies \citep{Deo2006}. The luminous HII Region `A'
located in the western spiral arm  has the size typical of
giant star-forming complexes in nearby spiral galaxies.  Infrared and UV observations \citep{MunozMarin2007,RothbergJoseph2004} confirm that Mrk 334 is a starburst galaxy.  This explains why in the ionization diagrams the part of the data points that belong to the nucleus lie in the domain corresponding to the ionization by young stars.

Such a powerful burst of star formation in a rather compact region produces a  hot gas superwind. Low-density gas heated by frequent supernova explosions forces its way through the dense and cold gas of the disk to form a wide-cone outflow in the direction perpendicular to the galaxy plane. Superwind
is usually most conspicuous in edge-on galaxies with the cone of hot gas fully open toward the observer. However, Mrk 334 has a less convenient orientation and the cone is seen projected against the bright disk of the galaxy and hence the conclusions about the presence of a superwind are to be based on the circumstantial evidences. Firstly, negative [OIII]-line velocities are observed toward the nucleus of the galaxy  (the base of the outflow) suggesting  outward motions of highly ionized gas  with velocities of $200-300\km$ perpendicular to the galactic plane. Also, the $150-180\km$  motions  are  observed in the NaD absorption line. The velocities are larger in the high-excited gas than in neutral medium, which is typical for galactic winds \citep{Veilleux2005}.

A second indication of the superwind is provided by the observed asymmetry of the X-ray flux distribution with respect to the nucleus.

Interaction-related processes become important as close to the nucleus as at the distances of $1-2$~kpc from it. We see their footprints as  Regions `B' and `C' and a system of tidal arcs and envelopes extending out
to galactocentric distances of $40$~kpc.  Fig.~\ref{{f13}} shows schematically  the inner region of the galaxy. An analysis of the kinematics of gas and stars led us
to conclude that the orbits of the debris of the disrupted companion lie outside the disk of Mrk 334 and
cross it at a considerable inclination. In the region of this cross-point we observe a cavern with  gas density lower than that of the ambient surrounding medium (Region `B')  which has formed as a result of the crossing of the disk by a dense stellar condensation. The rotation velocity  at this location is  $200-250\km$, implying that the fragments of the companion also move with the  velocities of the same order relative to the gaseous disk.  Also, we  find  evidence of a high  velocity collision in the gas kinematics and ionization. The collision has strongly perturbed the velocity field in the $\sim 3$ kpc size region. The maximum amplitude of the line-of-sight velocity perturbations amounts
to  $70\km$ in low-excitation lines and reaches $150\km$ in the  [OIII], because in this line we see low-density gas ionized by a powerful shock. The ionization state in Region `B' can be described in terms of the  shock+precursor model for a shock propagating at a speed of  $250-350\km$.
The mutual agreement of all the three estimates for the collision velocity   supports the adopted interpretation of the formation of Region  `B'.

Unfortunately, in the literature we have not found any detailed simulations  of the gaseous  disk response to the intrusion of a self-gravitating body whose mass is small compared to that of the
entire galaxy. As a close analogy, we can mention the paper by \citet{Levy2000}, that briefly analysed the
response of a gaseous disk of a galaxy to the crossing by a globular cluster. Even in such a relatively
small-scale collision the resulting shock propagates to at least five to six vertical disk scale heights.

We believe that the remnant of the companion galaxy now observed as Region `C' is the most likely candidate object to have punched the disk of Mrk 334 and produced there a cavern of hot gas. The line-of-sight velocities of the stellar population associated with Region `C' differ appreciably (by almost $100\km$) from the velocities of the old stellar population in the galactic disk. How is Region  `C' located with respect to the observer? Residual
velocities of ionized gas in the cavern (Region `B') are mostly negative. This means that the body that produced the cavern crossed the disk plane moving toward us and must now be located above the galaxy disk with respect to the observer (Fig.~\ref{{f13}}). Residual velocities of the young stellar population   in Region `C' are also negative implying that the nucleus of the companion traversed less than  a quarter of its orbit after the collision. In accordance with the rotation curve we  estimate the dynamical age of the cavern in Region `B' as $t\leq \cdot10^7$ yr. Therefore, we indeed deal with a recent collision and the cavern has not yet cooled down or condensed.

Our photometric analysis  of the tidal filaments yields a mass ratio for the interacting galaxies ranging from $1/5$ to $1/3$. The merger must have
occurred between two gas-rich galaxies. Indeed,  the total luminosity of nonaxisymmetric features in the
optical  images is  $25-30\%$ of the total luminosity of the galaxy.  \citet{Hopkins2008} suggest that  the `excess flux' is even greater in the K band where its contribution amounts to  $45\%$. According to \citet{Hopkins2008}, such a high percentage of the excess flux can be reproduced in the model of gas-rich galaxies merging, and the process results in the formation of a  LIRG galaxy, just as in the case of Mrk 334.

In Mrk 334 we appear to observe a transition from the  LIRG stage to the phase of nuclear activity.  The fact that the activity of the nucleus of the galaxy has started only recently is proved by weak water-ice absorptions in the infrared  spectra that are indicative of a certain well-defined phase in the evolution of the object \citep{Spoon2002}.

Thus, both the burst of star formation and nuclear activity in Mrk 334 date back to a rather recent epoch and their age  is comparable to the dynamic time scale of the interaction.   \citet{Li2008} pointed out, by references to Yuan et al.(in preparation), that ULIRGs experienced the stage of  `diffuse merger' when the nuclei of the interacting galaxies already merge together but have not yet formed a single nucleus. Composite activity --- active nucleus + star formation --- intensifies abruptly during this stage, and this appears to be now the case in  Mrk 334.

\begin{figure}
\includegraphics[width=8cm]{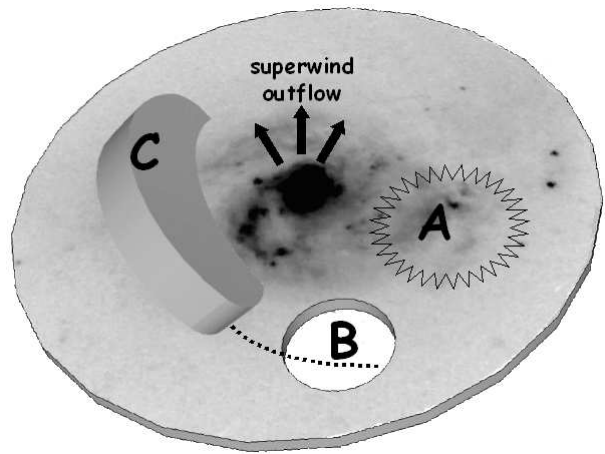}
\caption{Sketch of the proposed model describing the spatial structure of the inner ($r<5$ kpc) region of Mrk 334. The {\it HST} image from Fig.~\ref{f05} is projected onto the plane of  the galactic disk}
 \label{{f13}}
\end{figure}

\section{Conclusions.}
\label{sec7}

We used 3D spectroscopic data  and deep images   to explore the structure and kinematics of the galaxy Mrk 334. The galaxy has a composite (AGN+starburst) nucleus and  extended tidal structures in the form of loops and arcs observed at
galactocentric distances ranging from 2 to 40~kpc. Extensive spectroscopic and photometric data allowed
us to thoroughly analyse the structure of the inner regions of the galaxy. We consider the following
points to be the most important:

\begin{enumerate}
\item The main galaxy-to-companion mass ratio is about $3$ to $5$. The average surface-brightness profile shows a multi-tired structure and can be decomposed into a bulge and two exponential disks. We caught the galaxy in the process of the formation of an outer low-surface-brightness disk from the debris of the disrupted companion.

\item  The central region of the galaxy demonstrates a powerful starburst that must have
been triggered by the galaxy merger. Circumnuclear star formation is so intense that its contribution to the total ionization of
gas exceeds that of the active nucleus. As a result, the corresponding data points in the diagnostic
diagrams lie in the  HII/LINER domain. Such a powerful burst of star formation in a compact region gives rise to a  superwind with velocities of  $200-300\km$.   The asymmetric X-ray brightness distribution on \textit{ROSAT} maps is consistent with this  hypothesis.

\item We revealed a  region  $\sim2$~kpc east of the centre that is possibly the nucleus of a disrupted companion. The spectrum of this region exhibits stellar absorptions that are typical of a region that has undergone a burst of star formation about one Gyr ago. The radial velocities of the stars located in this region differ appreciably  from the stellar disk  of Mrk 334 onto which it is projected.   In the disk of the galaxy we found a cavern filled with low-density ionized gas. We interpret this region as a site of a recent  (about 10 Myr ago) crossing of the gaseous disk by the remnants of the disrupted companion. This supposition  allows us to explain
the unusually high [OIII]$/$H$\beta$ line ratio observed in this region that can be produced by a powerful shock propagating with a velocity of more than 250~$\km$. The non-circular gas motions  in this region   agree with  the crossing of the galaxy disk by the debris of the companion.
\end{enumerate}

This work is based on observations made with the 6-m telescope of the Special Astrophysical Observatory of the
Russian Academy of Sciences  operated under the financial support of the Ministry of Science of the
Russian Federation (Registration Number 01-43). This research has made use of the NASA/IPAC Extragalactic
Database (NED)  operated by the Jet Propulsion Laboratory, California Institute of Technology, under contract with the National Aeronautics and Space Administration. The research is partly based on data obtained from the Multimission Archive at the Space Telescope Science Institute (MAST). STScI is operated by the Association of Universities for Research in Astronomy, Inc., under NASA contract NAS5-26555.
We are grateful to Evgenii~Churazov for his assistance in discussing the X-ray data and to Olga Sil'chenko,  Victor Afanasiev and Natalia~Sotnikova for useful comments; and also to our anonymous referee for his constructive advice that helped us to improve the paper. This work was supported by the Russian Foundation for Basic Research (project no.~09-02-00870). AM is also grateful to   to the `Dynasty' Fund.

\end{document}